\documentclass{article}

\usepackage{arxiv}

\usepackage[utf8]{inputenc} 
\usepackage[T1]{fontenc}    
\usepackage{hyperref}       
\usepackage{url}            
\usepackage{booktabs}       
\usepackage{amsfonts}       
\usepackage{nicefrac}       
\usepackage{microtype}      
\usepackage{lipsum}		
\usepackage{graphicx}
\usepackage{doi}

\usepackage{times,epsfig,makeidx,amsfonts,amsmath,amsthm,dcolumn,multirow,graphicx,amssymb,colortbl,bm,url}
\usepackage{algorithm}
\usepackage[]{algorithmic}
\usepackage{afterpage,lscape}
\usepackage{paralist}

\usepackage{graphicx}
\usepackage{subfigure}
\usepackage{epstopdf}

\usepackage{mathrsfs}

\newtheorem{theorem}{Theorem}

\newcommand{\xin}{{\mbox{\boldmath $\xi$}}}

\newcommand{\bI}{{\bf I}}

\newcommand{\bt}{{\bf t}}
\newcommand{\bu}{{\bf u}}
\newcommand{\bx}{{\bf x}}

\newcommand{\bz}{{\bf z}}

\newcommand{\bX}{{\bf X}}

\newcommand{\bbeta}{\bm{\beta}}
\newcommand{\bmeta}{\bm{\eta}}
\newcommand{\balpha}{\bm{\alpha}}
\newcommand{\btheta}{\bm{\theta}}

\title{MEGH: A parametric class of general hazard models for clustered survival data }

\author{ \href{https://orcid.org/0000-0001-7183-8407}{\includegraphics[scale=0.06]{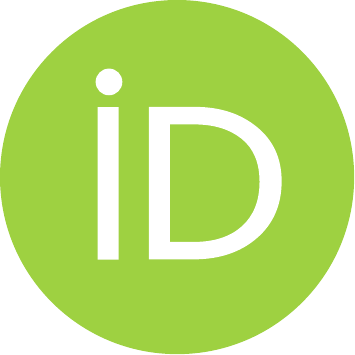}\hspace{1mm}Francisco Javier Rubio} \\
	Department of Statistical Science\\
	University College London \\
	London, UK\\
	\texttt{f.j.rubio@ucl.ac.uk} 
	\And
	\href{https://orcid.org/0000-0002-7245-9713}{\includegraphics[scale=0.06]{orcid.pdf}\hspace{1mm}Reza Drikvandi}\\
Department of Mathematical Sciences\\
Durham University\\
Durham, UK\\
	\texttt{reza.drikvandi@durham.ac.uk} \\
	}

\hypersetup{
pdftitle={XXX},
pdfsubject={stat.ME, stat.AP},
pdfauthor={Rubio, Drikvandi},
}

\begin{document}
\maketitle

\begin{abstract}
In many applications of survival data analysis, the individuals are treated in different medical centres or belong to different clusters defined by geographical or administrative regions. The analysis of such data requires accounting for between-cluster variability. Ignoring such variability would impose unrealistic assumptions in the analysis and could affect the inference on the statistical models. We develop a novel parametric mixed-effects general hazard (MEGH) model that is particularly suitable for the analysis of clustered survival data. The proposed structure generalises the mixed-effects proportional hazards (MEPH) and mixed-effects accelerated failure time (MEAFT) structures, among other structures, which are obtained as special cases of the MEGH structure.
We develop a likelihood-based algorithm for parameter estimation in general subclasses of the MEGH model, which is implemented in our R package {\tt MEGH}. We propose diagnostic tools for assessing the random effects and their distributional assumption in the proposed MEGH model. We investigate the performance of the MEGH model using theoretical and simulation studies, as well as a real data application on leukemia.
\end{abstract}



\section{Introduction}
In the analysis of time-to-event data, it is common to obtain clustered samples. For instance, in medical statistics, one may be interested in modelling the survival times of a sample of patients who receive attention in different health centres with varying characteristics or when the patients live in different geographic or administrative regions. This clustering or grouping information needs to be incorporated in any survival model to avoid under-estimation of the standard errors of the parameter estimators. Moreover, ignoring unobserved heterogeneity, due to clustering, in hazard regression models has an effect on model misspecification as the marginal model obtained by integrating out random effects may not coincide with the model ignoring clustering \cite{Struthers:1986,Omori:1993,Hougaard:1995}. A simple way to account for clustered survival data consists of fitting a stratified survival model. This is typically done by fitting a proportional hazards model for each strata level with different baseline hazards but sharing the same regression coefficients \cite{Collett:2015}. Alternatively, when the number of clusters is small, the cluster indicator can be coded as a categorical variable. The main disadvantage of these approaches is that they do not account for between-cluster variability. Modelling the clustering effect is more formally done by incorporating random effects, which are random variables that capture the hierarchical structure of the data and account for unobserved variability or heterogeneity between clusters. Regression models, including survival models, that incorporate random effects are often referred to as mixed-effects models (we refer the reader to \cite{mcculloch:2008} and the references therein for an extensive review of this kind of models). The most common mixed-effects models for time-to-event data correspond to extensions of common regression survival models. For example, the mixed-effects proportional hazards (MEPH) model \cite{therneau:2000,henderson:2002} is an extension of the semiparametric Cox proportional hazards models. This model is obtained by adding random effects, indexed by the cluster indicator, to the linear predictor function in the Cox model. Another common strategy to capture clustering consists of adding random effects to the linear predictor function in parametric and semiparametric  accelerated failure time (AFT) models, leading to the so-called mixed-effects AFT (MEAFT) model (see \cite{rubio:2018b} and \cite{do:2005} for a review on these models). A related class of mixed-effects models is constructed by adding random effects or Gaussian processes to the linear predictor function in hazard-based (parametric and semiparametric) regression models \cite{li:2002,austin:2017}. Some examples of these kinds of models are the mixed-effects spline regression models formulated at the cumulative hazard level by \cite{crowther:2014,crowther:2019,wood:2017}, or those formulated at the hazard level by \cite{vaida:2000} and \cite{charvat:2016}. In a related vein, \cite{zhou:2018} proposed mixed-effects models by adding random effects to proportional hazards, proportional odds, and accelerated failure time models in a Bayesian semiparametric framework. These classes of models are thus characterised by aiming at capturing the clustering effect via the incorporation of random effects into the linear predictor of different sorts of fixed effects hazard-based regression models. 

A natural question about mixed-effects models is the effect of misspecification of the random effects structure on the estimation of quantities of interest. It has been found that misspecifying the random-effects distribution has a negligible effect on point estimation of the fixed effects parameters of some classes of random effects models \cite{verbeke:1997}. However, there are more tangible effects of this misspecification on interval estimation of the parameters and predictions \cite{mcculloch:2011,rubio:2018b} as well as point and interval estimation of the parameters of the baseline hazard \cite{rossell:2019}. More crucially, ignoring a significant random effect could affect both parameter estimation and inference \cite{heagerty2001misspecified}. Tools for diagnosing misspecification of the random-effects part have been developed for mixed-effects models (without censoring) in recent years (see \cite{drikvandi:2017,efendi:2017,rao:2019}).

We develop a novel mixed-effects general hazard (MEGH) structure which incorporates random effects into a hazard-based regression model \cite{chen:2001,rubio:2019S} at the hazard scale and the time scale. The resulting model can capture clustering affecting the hazard scale and the time scale, and contains the MEPH and MEAFT models as particular cases. We focus our attention on two tractable subclasses of mixed-effects models that generalise the MEPH and MEAFT models.
The specification of the MEGH model requires modelling a baseline hazard, for which we employ several flexible parametric distributions. We provide diagnostic tools for the random-effects part of the proposed classes of models in the presence of censoring. We conduct extensive simulation studies that reveal that not only misspecifying the random-effects distribution is problematic, but also misspecifying the role of the random effects in the hazard structure has adverse effects.

The remainder of the paper is organised as follows. In Section \ref{sec_MEGH}, we describe the proposed MEGH model and discuss two submodels of interest. In Section \ref{sec:likelihood}, we present the expressions for the conditional and marginal likelihood functions, and define the marginal maximum likelihood estimator (MMLE), which defines the estimation approach followed in this paper. We also present theoretical results regarding the consistency and asymptotic normality of the MMLE under standard regularity conditions. Section \ref{sec:diagnostic} presents a test for the inclusion of random effects, as well as a graphical diagnostic tool to assess the goodness-of-fit of the random effects distribution. Section \ref{sec:simulation} presents an extensive simulation study which illustrates the performance of the proposed methodology as well as an exploration of the effects of misspecifying the distribution and the role of the random effects. Section \ref{realdatasection} presents an application with real data in the context of cancer survival with a spatial component. We conclude with a general discussion and possible extensions of our work in Section \ref{sec:discussion}.

\section{MEGH: Mixed-Effects General Hazard Models}
\label{sec_MEGH}
In this section, we describe the proposed MEGH model, and discuss two general subclasess of interest. To this end, let us first introduce some notation.
Let $o_{ij} \in {\mathbb R}_+$, $i=1,\dots,r$ indicates the cluster, $j=1,\dots,n_i$ denotes the individuals, be a sample of  times-to-event of interest, and $\bx_{ij}\in{\mathbb R}^p$ be a vector of covariates associated with the $ij$th individual. Let $c_{ij} \in {\mathbb R}_+$ be the right-censoring times, and $t_{ij} = \min\{o_{ij},c_{ij}\}$ be the observed survival times. Let $d_{ij} = I(o_{ij} < c_{ij})$ be the vital status indicators (1 - dead, 0 - alive), and $n = \sum_{i = 1}^r n_i$ be the total sample size across $r$ clusters.
We propose the mixed-effects general hazards (MEGH) model, which is defined through the individual hazard function:
\begin{equation}
\label{eq:MEGH}
h(t_{ij}\mid \bx_{ij}, u_i, \tilde{u}_i) = h_0\left(t_{ij} \exp\left\{\tilde{\bx}_{ij}^{\top}\balpha + \tilde{u}_i \right\}\right)\exp\left\{\bx_{ij}^{\top}\bbeta + u_i \right\},	
\end{equation}
where $h_0(\cdot)$ is a baseline hazard, $\bbeta=(\beta_1,\ldots,\beta_p)^\top \in \mathbb{R}^p$, $\balpha=(\alpha_1,\ldots,\alpha_{\tilde{p}})^\top \in \mathbb{R}^{\tilde{p}}$, 
${\bx}_{ij}\in \mathbb{R}^p$ is a vector of covariates affecting the hazard scale, $\tilde{\bx}_{ij} \in \mathbb{R}^{\tilde{p}}$ is a vector of covariates affecting the time scale (typically, $\tilde{\bx}_{ij} \subset \bx_{ij}$), and $(u_i,\tilde{u}_i) \stackrel{iid}{\sim} G$, where $G$ is a continuous distribution with support on ${\mathbb R}^{2}$ and zero mean.
Denote by  $\bX = \{\bx_{ij}\}$, $\tilde{\bX} = \{\tilde{\bx}_{ij}\}$,  the design matrices, and by $ \bX_o, \tilde{\bX}_o$ the  sub-matrices with the rows associated with the uncensored individuals. 
To avoid collinearity problems, we assume that the matrices associated with uncensored individuals are full column rank. 
The MEGH model represents an extension of the general hazard (GH) structure in \cite{chen:2001}, which is obtained by removing the random effects from \eqref{eq:MEGH}. 
Clearly, the GH structure contains the proportional hazards (PH) model for $\balpha=0$, the accelerated failure time (AFT) model for $\balpha=\bbeta$, and the accelerated hazards (AH) model for $\bbeta=0$ \cite{chen:2001}. 
Consequently, we also need to impose that the baseline hazard $h_0$ does not belong to the Weibull family, as in this case the GH structure, and consequently the MEGH structure, becomes non-identifiable \cite{chen:2001}.
Random effects are typically incorporated at the hazard scale (MEPH) or simultaneously at the hazard and time scales (MEAFT). The MEGH structure allows for incorporating random effects in both scales separately. The MEGH structure can also be seen as a generalisation of shared frailty survival models, which aim at accounting for unobserved heterogeneity between clusters \cite{wienke:2010,charvat:2016}. 
Another appealing feature of the MEGH formulation is that the cumulative hazard function can be written in closed-form as follows
\begin{equation*}
H(t_{ij}\mid \bx_{ij}, u_i, \tilde{u}_i) = H_0\left(t_{ij} \exp\left\{\tilde{\bx}_{ij}^{\top}\balpha+ \tilde{u}_i \right\}\right)\exp\left\{\bx_{ij}^{\top}\bbeta - \tilde{\bx}_{ij}^{\top}\balpha + u_i -  \tilde{u}_i\right\}.	
\end{equation*}
The MEGH structure \eqref{eq:MEGH} is a very rich mixed hazard structure that contains, as particular cases, a number of hazard models of interest. In this paper, we particularly focus on the following two subclasses.


\subsection*{MEGH-I: Mixed Structure I}
The first subclass, MEGH-I, is defined by the hazard and cumulative hazard functions:
\begin{equation}
\begin{split}
h_1(t_{ij}\mid \bx_{ij}, u_i) &= h_0\left(t_{ij} \exp\left\{\tilde{\bx}_{ij}^{\top}\balpha\right\}\right)\exp\left\{\bx_{ij}^{\top}\bbeta + u_i \right\},	\\
H_1(t_{ij}\mid \bx_{ij}, u_i) &= H_0\left(t_{ij} \exp\left\{\tilde{\bx}_{ij}^{\top}\balpha\right\}\right)\exp\left\{\bx_{ij}^{\top}\bbeta - \tilde{\bx}_{ij}^{\top}\balpha + u_i \right\}.
\end{split}
\label{eq:MEGH1}	
\end{equation}
The mixed hazard structure \eqref{eq:MEGH1} contains the MEPH model ($\balpha = 0$) \cite{therneau:2000}, while allowing for the inclusion of time-dependent effects through the parameter $\balpha$.  

\subsection*{MEGH-II: Mixed Structure II}
The second subclass, MEGH-II, is defined by the hazard and cumulative hazard functions:
\begin{equation}
\begin{split}
h_2(t_{ij}\mid \bx_{ij}, u_i) &= h_0\left(t_{ij} \exp\left\{\tilde{\bx}_{ij}^{\top}\balpha + u_i \right\}\right)\exp\left\{\bx_{ij}^{\top}\bbeta + u_i \right\},	\\
H_2(t_{ij}\mid \bx_{ij}, u_i) &= H_0\left(t_{ij} \exp\left\{\tilde{\bx}_{ij}^{\top}\balpha+ u_i \right\}\right)\exp\left\{\bx_{ij}^{\top}\bbeta - \tilde{\bx}_{ij}^{\top}\balpha  \right\},	
\end{split}
\label{eq:MEGH2}	
\end{equation}
The MEGH-II structure \eqref{eq:MEGH2} generalises the MEAFT structure ($\balpha=\bbeta$).

\begin{figure}
\centering
\caption{Diagram of the link between the MEGH and some particular hazard structures of interest. (AH =  accelerated hazards, AFT= accelerated failure time, PH = proportional hazards, GH = general hazard, MEPH = mixed effects proportional hazards, MEAFT = mixed effects accelerated failure time, MEGH = mixed effects general hazard)}
\label{fig:diag}
\includegraphics[scale=1]{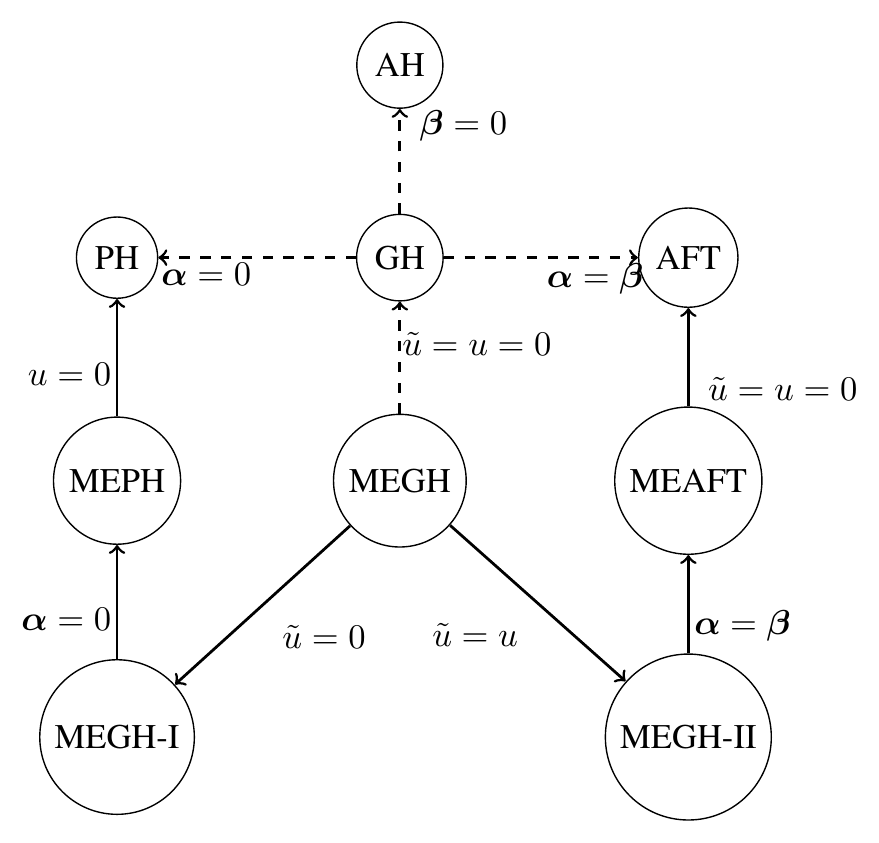}
\end{figure}

Figure \ref{fig:diag} shows the link between the MEGH and some submodels of interest, while the proposed MEGH structure is much richer than illustrated in this figure.

We will adopt parametric baseline hazards $h_0(\cdot \mid \btheta)$ for models \eqref{eq:MEGH1}-\eqref{eq:MEGH2}, based on flexible parametric distributions. This is an important step as the choice of the parametric baseline hazard determines the hazard shapes one can obtain. For instance, the log-normal hazard function is unimodal (up-then-down), while the Gamma hazard function can be increasing, decreasing or flat. There exists other (three-parameter) distributions that can capture the basic shapes of the hazard (increasing, decreasing, unimodal, and bathtub), such as the Exponentiated Weibull, Generalised Gamma and Power Generalised Weibull (PGW). However, it is important to keep in mind that an efficient estimation of the parameters of more flexible distributions typically requires larger sample sizes \cite{rubio:2019S,alvares:2021}. On the other hand, parametric models have been found to perform well, compared to semiparametric counterparts, in the context of clustered survival data \cite{prenen:2017}.

\section{Likelihood Function and Parameter Estimation}\label{sec:likelihood}
We here describe the conditional and marginal likelihood functions, which will be used to calculate the marginal maximum likelihood estimators as well as to develop the diagnostic tools presented in the next section. We discuss the details behind the calculation of the marginal likelihoods and the optimisation methods. Finally, we present a theoretical result on the consistency and asymptotic normality of the marginal maximum likelihood estimators.

\subsection{Likelihood and Marginal Likelihood Functions}
Let $\bt_i = (t_{i1}, \dots, t_{i{n_i}})$ be the sample associated with the $i$th cluster and $\bx_i = (\bx_{i1},\dots, \bx_{i{n_i}})$ denote the corresponding matrix of covariates. 
The cluster-specific marginal likelihood function associated with the $i$th cluster, after integrating out the random effects $u_i$ and $\tilde{u}_i$, is given by
$$ \int_{\mathbb{R}^{2}} \prod_{j=1}^{n_i} h(t_{ij}\mid \bx_{ij}, u_i, \tilde{u}_i)^{d_{ij}} \exp\left\{ - H(t_{ij}\mid \bx_{ij}, u_i, \tilde{u}_i)  \right\}  dG(u_i,\tilde{u}_i).$$
For the case where the random-effects distribution $G$ is assumed to be a multivariate distribution with mean $\boldsymbol{0}$ and finite variance, with parameters $\xin$,
we can write down the marginal likelihood function of the MEGH model \eqref{eq:MEGH} in terms of the hazard and cumulative hazard functions as follows. To simplify notation, let us denote $\bmeta = (\bbeta,\balpha, \btheta,\xin)$ and let $\Gamma$ be the parameter space. The marginal likelihood function is
\begin{equation}\label{eq:marglik}
m(\bmeta ) = \prod_{i=1}^{r} m_{i}(\bmeta),
\end{equation}
where $ m_{i}$ represents the cluster-specific marginal likelihood associated with the $i$th cluster
\begin{equation}\label{eq:clustmarglik}
m_{i}(\bmeta) =
\int_{\mathbb{R}^{2}}  \exp\left\{ \ell_{i}(\bmeta,u_i, \tilde{u}_i )\right\}dG(u_i,\tilde{u}_i;\xin),
\end{equation}
and
\begin{equation*}
\ell_{i}(\bmeta,u_i, \tilde{u}_i ) = \sum_{j=1}^{n_i} {d_{ij}}\log h(t_{ij}\mid \bx_{ij},u_i, \tilde{u}_i ) - \sum_{j=1}^{n_i} H(t_{ij}\mid \bx_{ij},u_i, \tilde{u}_i ),
\end{equation*}
denotes the log-likelihood function conditional on the random effects $u_i$ and $\tilde{u}_i$. The marginal maximum likelihood estimator (MMLE) is then defined as $\widehat{\bmeta} = \operatorname{argmax}_{\Gamma} m(\bmeta)$.

\subsection{Computations}
To calculate the marginal maximum likelihood estimator $\widehat{\bmeta}$, we need to evaluate the marginal likelihood function \eqref{eq:marglik}, which is a product of integrals that define the cluster-specific marginal likelihoods \eqref{eq:clustmarglik}. The integrand in \eqref{eq:clustmarglik} can be small when the number of individuals belonging to that cluster is moderate or large. To deal with this numerical challenge, we scale the integrand by using 
\begin{equation*}
m_{i}(\bmeta) = K_i
\int_{\mathbb{R}^{2}} \exp\left\{ \ell_{i}(\bmeta,u_i, \tilde{u}_i ) - \log K_i \right\}dG(u_i,\tilde{u}_i),
\end{equation*}
where $\log K_i = \max_{u_i, \tilde{u}_i} \ell_{i}(\bmeta,u_i, \tilde{u}_i )$, for fixed values of $\bmeta$. This allows for a more stable numerical integration, as the integrand is now bounded between $(0,1]$ (always taking the value $1$ at the maximum). Maximisation over the random effects is simpler as their variance is rarely large. Numerical integration in our empirical examples, which include only one hierarchical level, is done by using the R command \texttt{integrate} with default options. Nonetheless, other methods, such as Laplace approximations or Monte Carlo integration, could also be used. The optimisation step is carried out using standard R routines, such as those implemented in the commands {\tt optim} and {\tt nlminb}. The R package {\tt MEGH} (\url{https://github.com/FJRubio67/MEGH}) implements the MEGH-I and MEGH-II models under these specifications.

\subsection{Theory}
Here, we prove the consistency and asymptotic normality of the MMLE, when $r\to \infty$ (which implicitly implies $n\to \infty$), under assumptions A1--A7 given in the online Supplementary Material. We denote by $\bmeta^{\star} = (\bbeta^{\star},\balpha^{\star}, \btheta^{\star},\xin^{\star})$ the unknown true value of the parameters, which is assumed to be an interior point of the parameter space $\Gamma$. 
%
\begin{theorem} \label{th:asymp}
	Consider the MEGH model \eqref{eq:MEGH}. Let $h_0\left(\cdot \mid \btheta \right)$ be a pre-specified parametric baseline hazard as in Section \ref{sec_MEGH}.
	Under assumptions A1--A7 given in the online Supplementary Material, it follows that 
	\begin{itemize}
		\item[(i)] Consistency: $\widehat{\bmeta} \stackrel{P}{\to} \bmeta^{\star}$ as $r\to \infty$.
		\item[(ii)] Asymptotic normality: $\sqrt{r}\left(\widehat{\bmeta} -  \bmeta^{\star} \right) \stackrel{d}{\to} N({\bf 0},\bI(\bmeta^{\star})^{-1})$ as $r\to \infty$,
	\end{itemize}
	where $\bI(\bmeta^{\star})$ is the expectation matrix $ \bI(\bmeta) = \operatorname{cov}\left[\nabla_{\bmeta} \log m_1(\bmeta) \right]$ evaluated at $\bmeta^{\star}$.
\end{theorem}
Regarding the required assumptions A1--A7, we note that assumption A1 is a restriction to a compact parameter space, which can be assumed to be a large compact set. 
Assumption A2 concerns the censoring mechanism, which is assumed to be non-informative.
Assumption A3 guarantees identifiability and continuity of the MEGH model. 
Assumption A4 togheter with assumptions A5--A7 represent standard integrability and differentiability conditions on the marginal likelihood function, which implicitly impose conditions on the choice of the baseline hazard. 
Assumption A6 assumes the existence and non-singularity of the  information matrix. Verifying such condition in practice for specific choices of the parametric baseline hazard is complicated. However, it has been shown that certain smoothness and identifiability conditions, which are easier to verify in practice, guarantee the non-singularity of this matrix (see \cite{Asgharian:2004}).
Similar asymptotic results have also been obtained for parametric copula survival models for clustered data in \cite{prenen:2017}, where parametric models were found to perform better than their semiparametric counterparts.

\section{Diagnostic Tools for Random Effects in the MEGH Model}\label{sec:diagnostic}
As already discussed, the proposed MEGH model \eqref{eq:MEGH} incorporates random effects to capture the unknown variability between clusters or groups in the survival data. Since random effects are latent and unobservable variables, their assumed distribution can be subject to misspecification, so it is important to check their distributional assumption to identify potential model misspecification. It is also important to test which random effects are statistically significant and should be included in the model. These two problems are crucial especially for practical use, so we study them for the MEGH model in the following two subsections respectively.

\subsection{Detecting Misspecification of the Random-Effects Distribution}
We here present a diagnostic tool for detecting misspecification of the random-effects distribution in the MEGH model. The assumed random-effects distribution $G$ is typically a multivariate normal distribution. To check the appropriateness of an assumed random-effects distribution, \cite{verbeke2013} suggested to use the so-called gradient function for models with random effects. Their diagnostic tool based on a gradient function is not directly applicable to the MEGH model  \eqref{eq:MEGH}. We extend the gradient function approach for the MEGH model. For this purpose, we use the key idea of the gradient function, that is, to check if the marginal log-likelihood of the model can be increased considerably by replacing $G$ with another random-effects distribution, say $W$. If so, then the assumed distribution $G$ may not be adequate for the random effects because another random-effects distribution produces a considerably larger marginal likelihood.

Recall that ${\boldsymbol{\eta}}=(\bbeta,\balpha, \btheta, \xin)$ is the combined vector of the model parameters. We here write the marginal likelihood as $m(G)$, instead of $m({\boldsymbol{\eta}})$, to emphasise its dependence on the assumed random-effects distribution $G$. To check if the marginal log-likelihood can be increased considerably by replacing $G$ with another random-effects distribution $W$, we use the directional derivative of the marginal log-likelihood $\log m(.)$ at $G$ into the direction $W$ as follows
\begin{eqnarray*}
	\Phi(G,W) &=& \mathop {\lim }\limits_{\varepsilon \mathop  \to \limits^ >  0} \frac{{\log m\big((1 - \varepsilon )G + \varepsilon W\big) - \log m(G)}}{\varepsilon } \\
	&=& \frac{{\partial \log m\big((1 - \varepsilon )G + \varepsilon W\big)}}{{\partial \varepsilon }}\mathop \big|\nolimits_{\varepsilon = 0}.
\end{eqnarray*}
Then, there is no better random-effects distribution than $G$ if $\Phi(G,W) \le 0$ for all $W$. We can write that
\begin{eqnarray*}
	\frac{1}{r}\Phi (G,W) &=& \frac{1}{r} \frac{{ \partial \sum_{i=1}^{r}{ \log [(1 - \varepsilon )m_i(G) + \varepsilon m_i(W)]}}}{{\partial \varepsilon }}\mathop \big|\nolimits_{\varepsilon = 0} \\
	&=& \frac{1}{r} \sum_{i=1}^{r}{ {\frac{{m_i(W) - m_i(G)}}{{m_i(G)}}}} \\
	&=& \frac{1}{r} \sum_{i=1}^{r}{{\frac{{\int_{\mathbb{R}^{2}} \exp\{\ell_{i}(u, \tilde{u} )\}dW(u,\tilde{u}) }}{{m_i(G)}}} - 1} \\
	&=& \int_{\mathbb{R}^{2}} \Big[\frac{1}{r} \sum_{i=1}^{r}{{\frac{\exp\{\ell_{i}(u, \tilde{u} )\}}{{m_i(G)}}}}\Big] dW(u,\tilde{u}) - 1 \\
	&=& \int_{\mathbb{R}^{2}} {\Delta \big(G,(u,\tilde{u}) \big)}dW(u,\tilde{u}) - 1,
\end{eqnarray*}
in which $\Delta \big(G,(u,\tilde{u}) \big)$ is the gradient function for the MEGH model, defined as
\begin{equation}
\label{grdint}
\Delta \big(G,(u,\tilde{u}) \big) = \frac{1}{r} \sum_{i=1}^{r}{{\frac{\exp\{\ell_{i}(u, \tilde{u} )\}}{{m_i(G)}}}}, \,\,\,\,\,\,\,\,\,\, (u,\tilde{u}) \in \mathbb{R}^{2}.
\end{equation}
Hence, we have $\phi(G,W) \le 0$ for all $W$, if $\Delta \big(G,(u,\tilde{u})\big) \le 1\,\,\forall\,(u,\tilde{u}) \in \mathbb{R}^{2}$. The gradient function (\ref{grdint}) has useful properties. The calculation of the gradient function does not require specifying an alternative distribution $W$. Also, since $\Phi(G,G)=0$, we must have $\Delta \big(G,(u,\tilde{u}) \big)=1$ for all random effects points $(u,\tilde{u})$ in the support of $G$. As a graphical diagnostic tool, we plot the gradient function versus random effects points $(u,\tilde{u})$, and if the gradient plot does not exceed $1$ then the assumed random-effects distribution $G$ will be deemed to be adequate for random effects; otherwise, the random-effects distribution is deemed to be misspecified. See \cite{verbeke2013} and \cite{Drikvandi:2020} who applied a similar approach to generalised linear and nonlinear mixed models. The calculation of the gradient function (\ref{grdint}) is straightforward as it only requires the calculation of the marginal and conditional distributions for all $r$ clusters. Note that the dependence of the gradient function on the parameters ${\boldsymbol{\eta}}=(\bbeta,\balpha, \btheta, \xin)$ is suppressed for simplicity of the notation, however we replace these parameters with their estimates when calculating the gradient function.

\subsection{Tests for the Need of Random Effects in the MEGH Model}
\label{sec_testsRE}
Since random effects are latent and unobservable variables, it is not obvious in practice which random effects must be included in the model. As an initial check, one can use a plot of the survival curves for all clusters to get some understanding of the variability between clusters and the potential need of random effects (see an example in Figure \ref{fig:KMs}). We provide formal tests for verifying which random effects are statistically significant and must be present in the MEGH model. If the test confirms that all the random effects are not significant, then the extended model \eqref{eq:MEGH} will not be essential and one could use the reduced models without random effects as in \cite{chen:2001} and \cite{rubio:2019S}.

The test for inclusion or exclusion of random effects can be carried out via a test for zero random effects. From a statistical perspective, to test for zero random effects is equivalent to testing if their variance parameters are equal to $0$.
We want to test whether or not the random effects $u_i$ and $\tilde{u}_i$ can be excluded from the MEGH model \eqref{eq:MEGH}. This hypothesis testing problem can be expressed as follows
\begin{equation}
	\label{test_sigma_u}
	\begin{cases}
		H_{0}: \sigma ^2_u = 0 \,\, {\text{and}} \,\, \sigma ^2_{\tilde{u}} = 0 \\
		H_{1}: \sigma ^2_u > 0 \,\, {\text{or}} \,\, \sigma ^2_{\tilde{u}} > 0.
	\end{cases}
\end{equation}

This is a non-standard testing problem because the null hypothesis places the true variance parameters on the boundary of the parameter space \cite{Self_Liang_1987,Stram_Lee_1994,Crainiceanu_Ruppert_JRSSB2004,Drikvandi_2013}. Classical tests such as the likelihood ratio, Wald and score tests suffer from testing on the boundary of the parameter space, because the regularity conditions do not hold under such situations. As a consequence, the usual asymptotic chi-squared distribution of the likelihood ratio or score statistic is incorrect here. The correct asymptotic distribution is well understood to be a mixture of chi-squared distributions when $r\to \infty$ under some mild conditions \cite{Self_Liang_1987,Stram_Lee_1994}. 

The hypotheses $H_0$ and $H_1$ in (\ref{test_sigma_u}) correspond to Case 7 of \cite{Self_Liang_1987}. Thus, under $H_0$, the correct asymptotic distribution of the likelihood ratio statistic for testing (\ref{test_sigma_u}) would be $\frac{1}{4}\chi _0 ^2  + \frac{1}{2}\chi _1 + \frac{1}{4}\chi _2 ^2$, where $\chi _0 ^2$ is a point mass at $0$, and $\chi _1 ^2$ and $\chi _2 ^2$ are the chi-squared distributions with one and two degrees of freedom, respectively. The $p$-value of the likelihood ratio test for the hypothesis test (\ref{test_sigma_u}) is as follows (see also \cite{Verbeke:2003})
\begin{equation}
\label{pvalue_sigma_u}
p = \frac{1}{4}P(\chi _0  \ge  R_{\text{obs}}) + \frac{1}{2}P(\chi _1  \ge  R_{\text{obs}}) + \frac{1}{4}P(\chi _2  \ge  R_{\text{obs}}),
\end{equation}
where $R_{\text{obs}}$ denotes the observed likelihood ratio statistic. 

For the MEGH model with the mixed hazard structures \eqref{eq:MEGH1}-\eqref{eq:MEGH2}, since $u_i$ and $\tilde{u}_i$ are the same, the hypothesis test (\ref{test_sigma_u}) collapses to
\begin{equation*}
\begin{cases}
H_{0}: \sigma ^2_u = 0 \\
H_{1}: \sigma ^2_u > 0.
\end{cases}
\end{equation*}
The correct asymptotic distribution of the likelihood ratio statistic for this hypothesis test is $\frac{1}{2}\chi _0 ^2  + \frac{1}{2}\chi _1$, as it corresponds to Case 5 of \cite{Self_Liang_1987}. 

We will apply the above diagnostic tools to our case study in Section \ref{realdatasection}. We note that there are also non-asymptotic tests available for testing random effects, which avoid the boundary issue. Bootstrap and permutation tests as well as Bayesian tests do not rely on asymptotic results and hence avoid boundary issues when testing random effects. Those tests have been well studied for linear and generalised linear mixed models \cite{Drikvandi_2013,Saville_2009,rao:2019}; however, such tests may not be easily applied to the case of MEGH model, due to the presence of censored and clustered data which makes bootstrap or permutation challenging. Furthermore, \cite{Verbeke:2003} and \cite{Claeskens:2008} provide asymptotic score tests that can be applied in this context.
\section{Simulation Studies }\label{sec:simulation}
In this section, we conduct simulations to evaluate the performance of the proposed MEGH model. In the simulations, we first examine the accuracy of parameter estimation as well as statistical inference with the MEGH model under different scenarios. We investigate how the MEGH model compares with the model ignoring random effects. Such comparisons are crucial in demonstrating the advantages of the proposed MEGH model over the models ignoring random effects. We then assess the impact of misspecifying the mixed structure as well as misspecifying the random-effects distribution. We also examine the finite-sample performance of the diagnostic tests for testing random effects in the MEGH model. Some of the simulation results are reported in the online supplementary material for the sake of space. 

In the simulations we study both mixed hazard structures \eqref{eq:MEGH1}-\eqref{eq:MEGH2} and consider the following MEGH model, according to our real data application in Section \ref{realdatasection}, 
\begin{equation}
\label{s1}
h(t_{ij}\mid \bx_{ij}, u_i, \tilde{u}_i) = h_0\left(t_{ij} \exp\left\{ \alpha_1{\text{age}}_{ij} + \tilde{u}_i \right\}\right)\exp\left\{ \beta_1{\text{age}}_{ij} + \beta_2{\text{sex}}_{ij} + \beta_3{\text{wbc}}_{ij} + \beta_4{\text{tpi}}_{ij} + u_i \right\}, 
\end{equation}
where the covariates $\bx_{ij}^{\top}=({\text{age}}_{ij},{\text{sex}}_{ij},{\text{wbc}}_{ij},{\text{tpi}}_{ij})$ are in accordance with the leukemia data in Section \ref{realdatasection}, and $h_0(\cdot)$ is the baseline hazard for which we here consider the PGW baseline hazard with parameters $\btheta=(\eta,\nu,\delta)$ as in the online supplementary material. We also conduct simulations with the log-logistic baseline hazard, which are presented in the supplementary material.

Recall that $\tilde{u}_i=0$ for the mixed hazard structure MEGH-I, and $\tilde{u}_i=u_i$ for the mixed hazard structure MEGH-II. In the simulations, we choose the true parameter values according to the estimates obtained from the leukemia data application. The number of clusters is 24. We simulate 250 data sets from the above model with each of the hazard structures MEGH-I and MEGH-II, each with the same size as of the real data application (\textit{i.e.}, $n=1043$), using our R package {\tt MEGH} available in the online supplementary material. The random effects are generated from a normal distribution with mean $0$ and two standard deviation values of $\sigma_u=0.5$ and $\sigma_u=1$. The censoring rate is set to $25\%$. 

\begin{figure}
	\centering
	\caption{The bias of the estimates from the three methods: MEGH-I, MEGH-II and MLE, for all the parameters based on $250$ simulation replications when the simulated data are generated from model (\ref{s1}) with the mixed structure I and PGW baseline hazard, and a normal distribution for the generated random effects with $\sigma_u=1$.}
	\label{PGW-fig1}
	\vspace*{2mm}
	\subfigure[Bias in estimating $\eta$]{\includegraphics[scale=0.17]{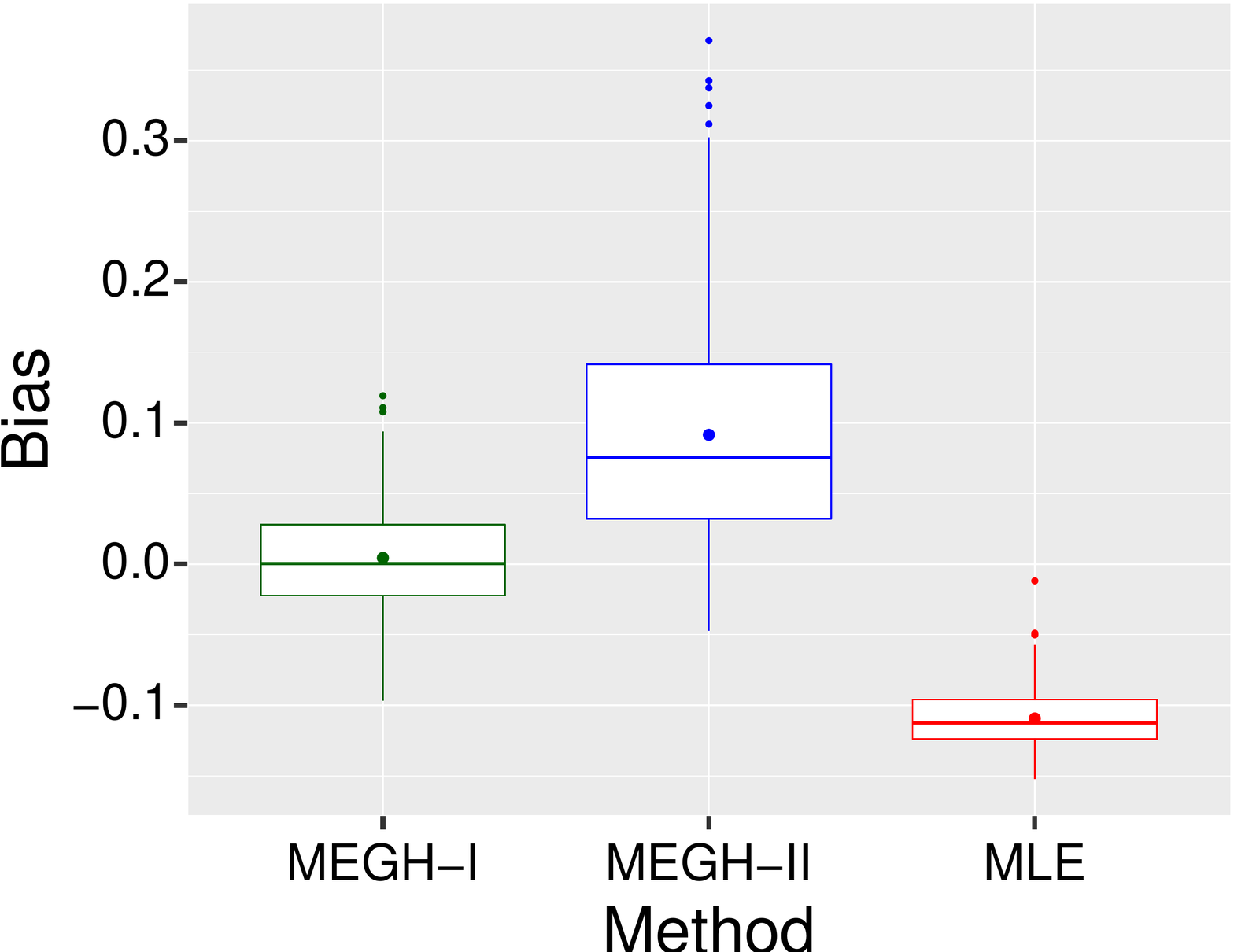}}
	\hspace{0.6cm}
	\subfigure[Bias in estimating $\nu$]{\includegraphics[scale=0.17]{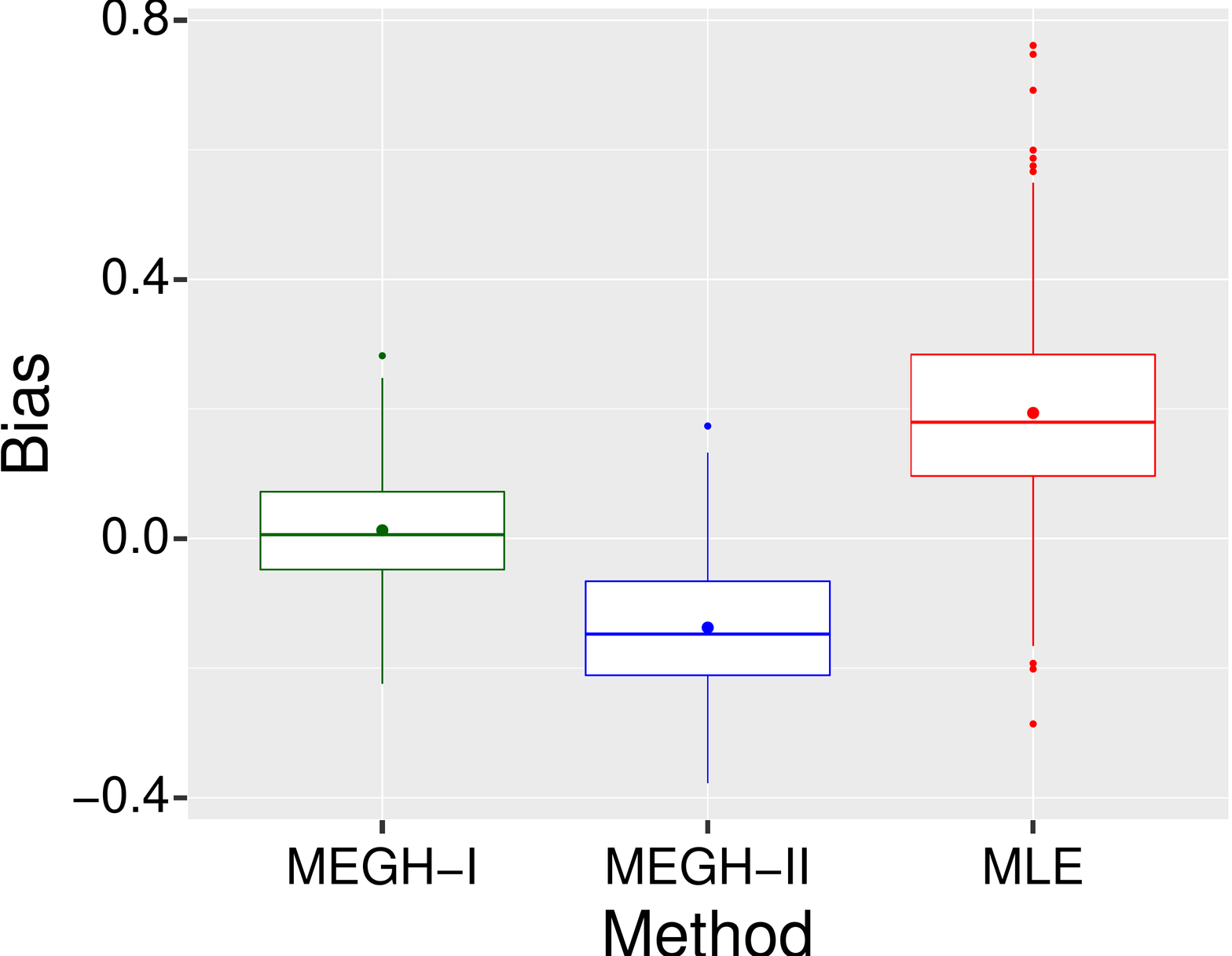}} 
	\hspace{0.6cm}
	\subfigure[Bias in estimating $\delta$]{\includegraphics[scale=0.17]{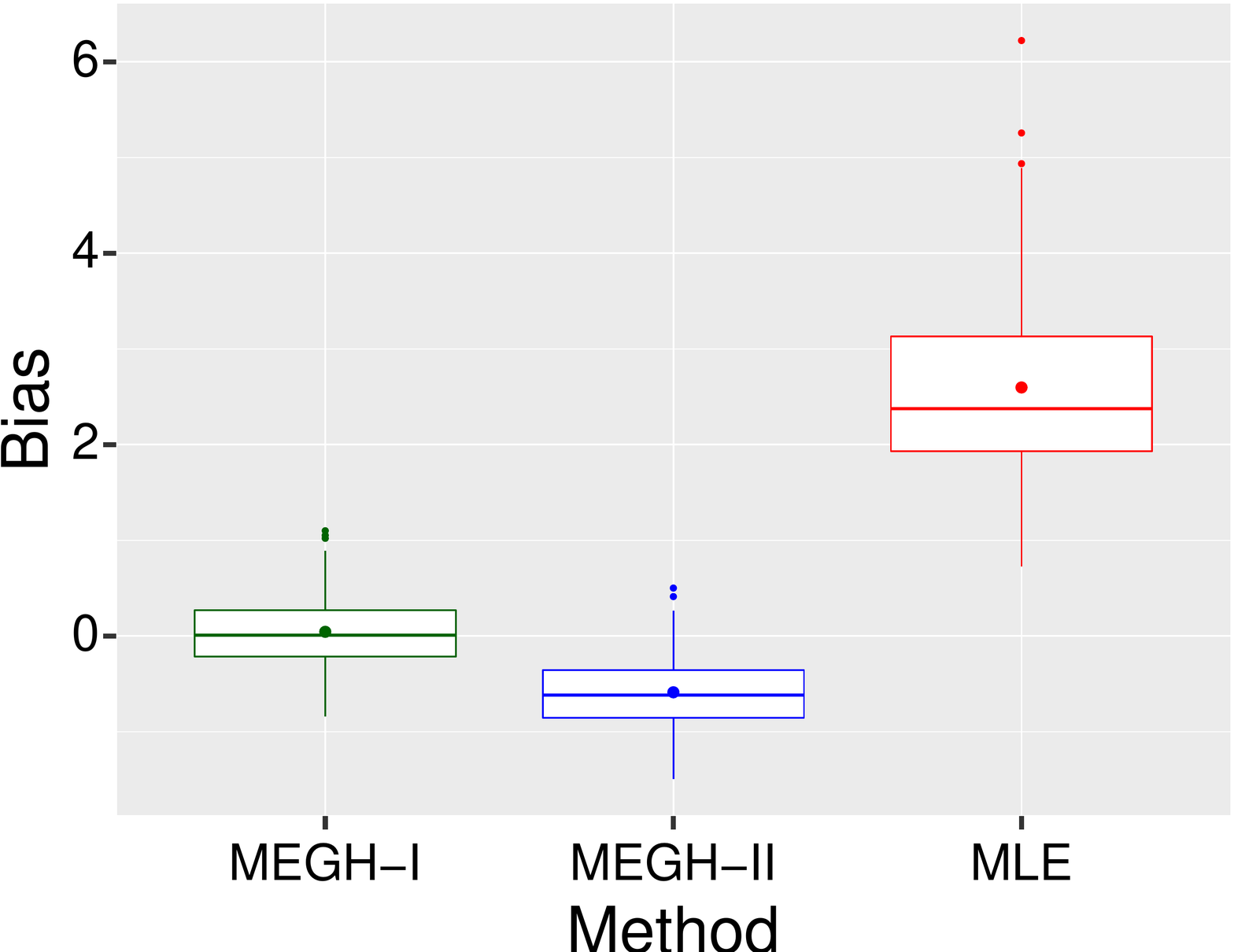}} \\
	\subfigure[Bias in estimating $\alpha$]{\includegraphics[scale=0.17]{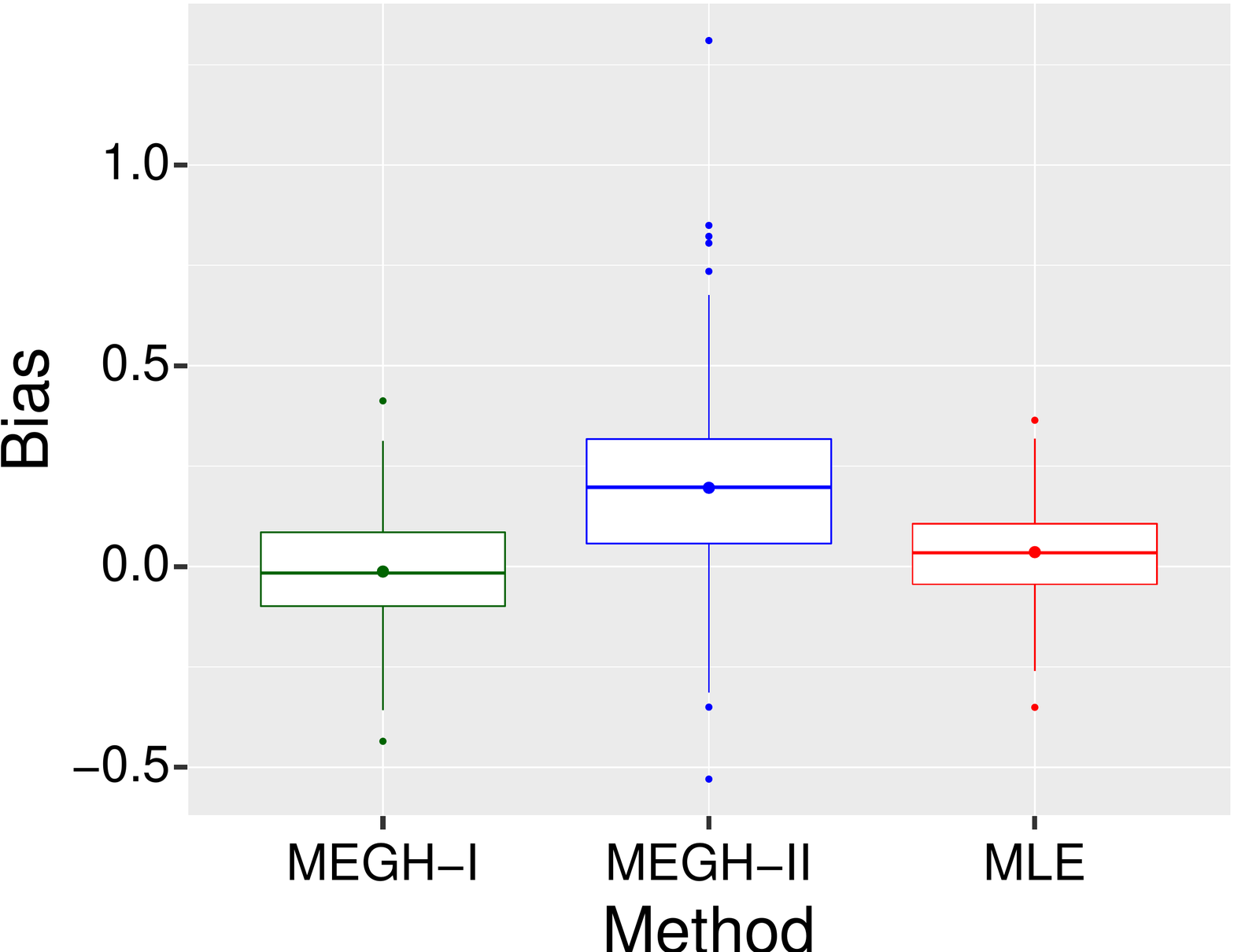}} 
	\hspace{0.6cm}
	\subfigure[Bias in estimating $\beta_1$]{\includegraphics[scale=0.17]{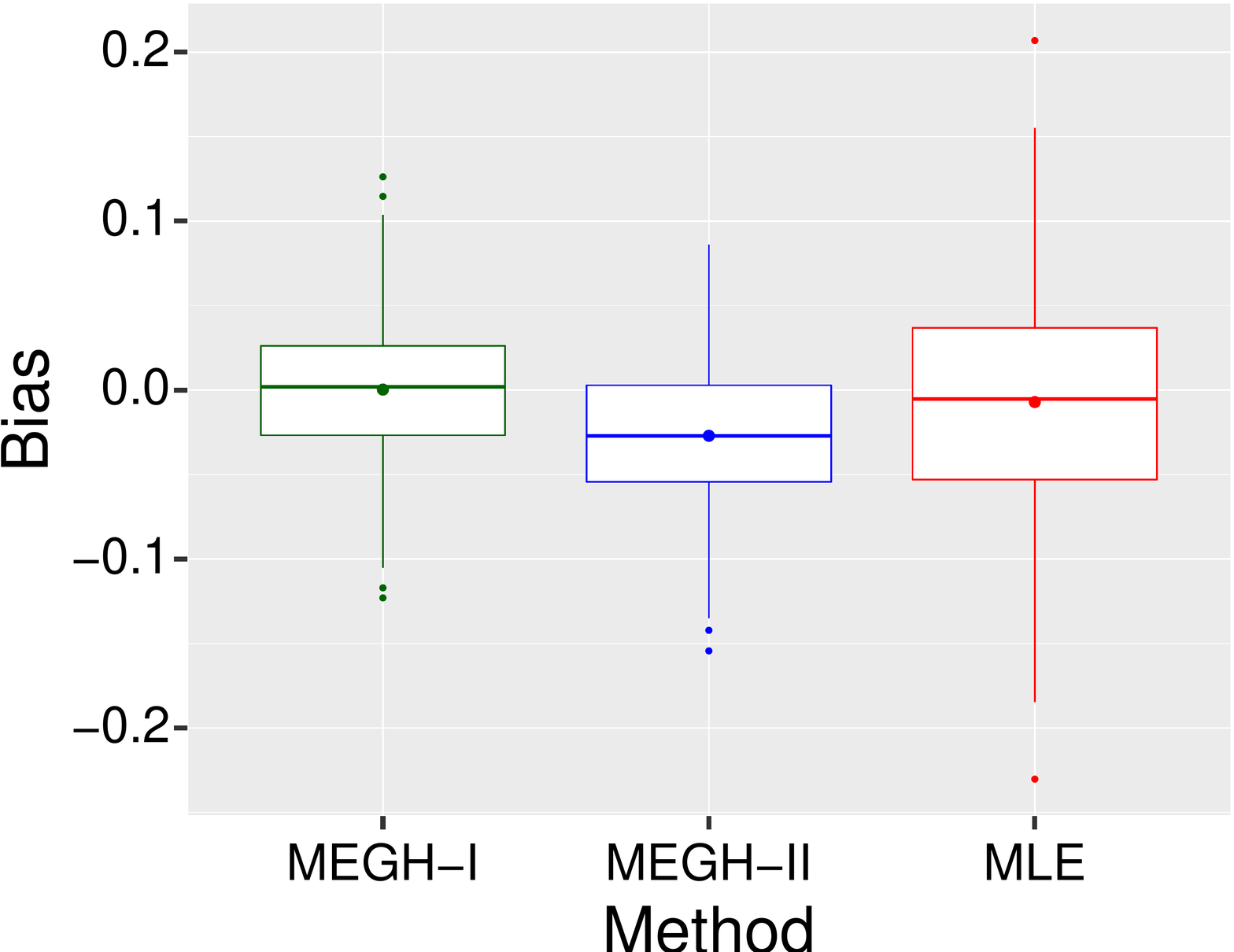}} 
	\hspace{0.6cm}
	\subfigure[Bias in estimating $\beta_2$]{\includegraphics[scale=0.17]{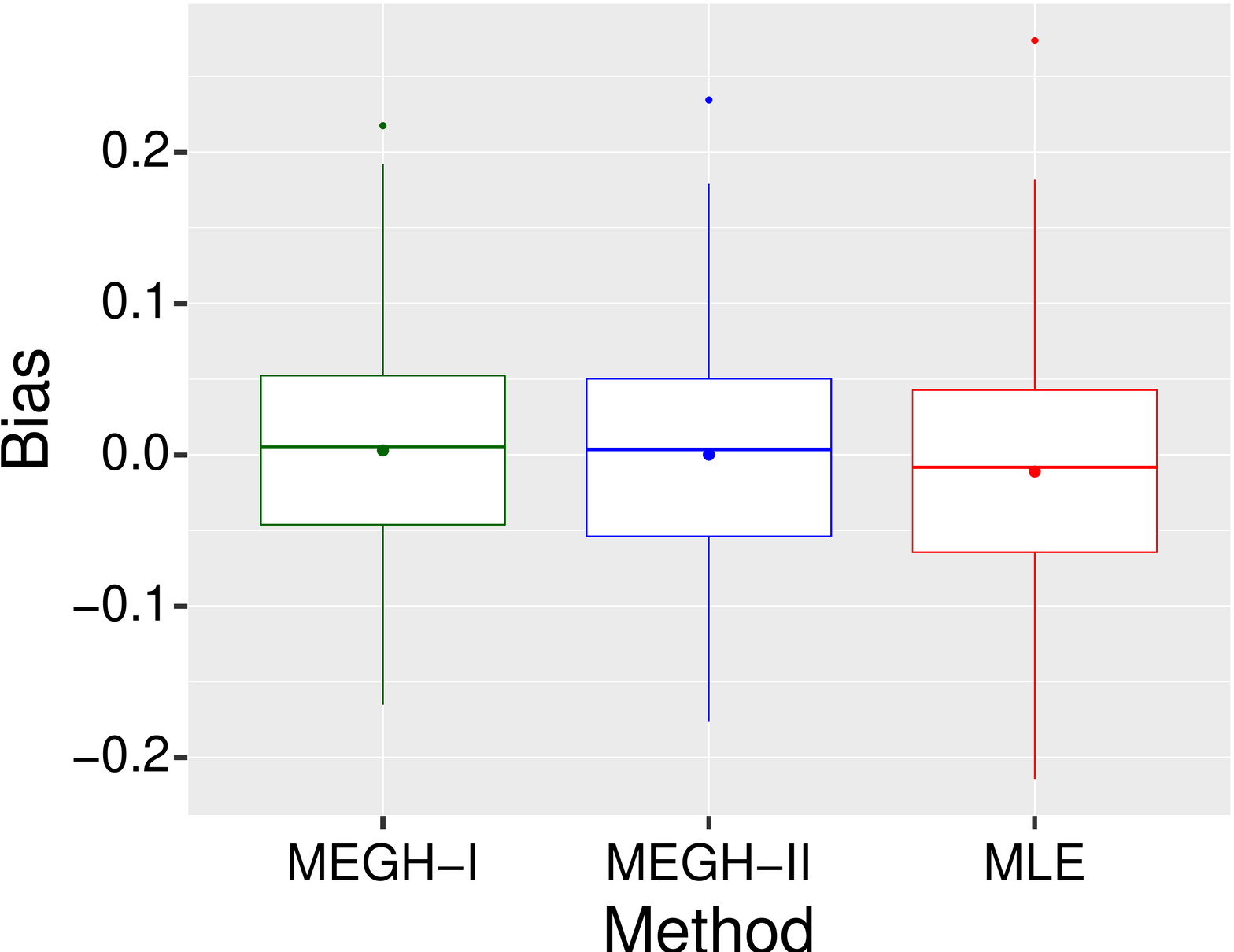}} \\
	\subfigure[Bias in estimating $\beta_3$]{\includegraphics[scale=0.17]{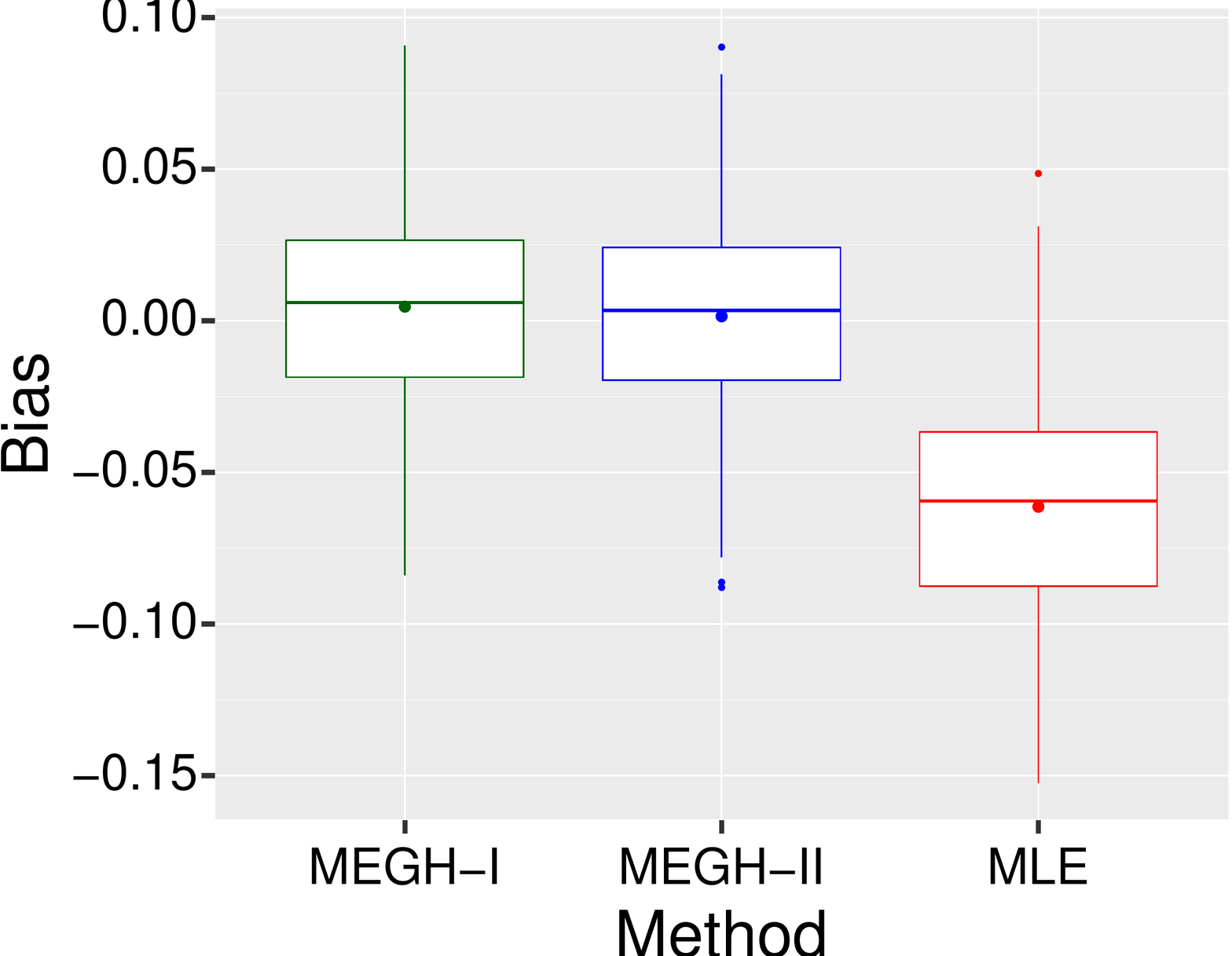}}
	\hspace{0.6cm}
	\subfigure[Bias in estimating $\beta_4$]{\includegraphics[scale=0.17]{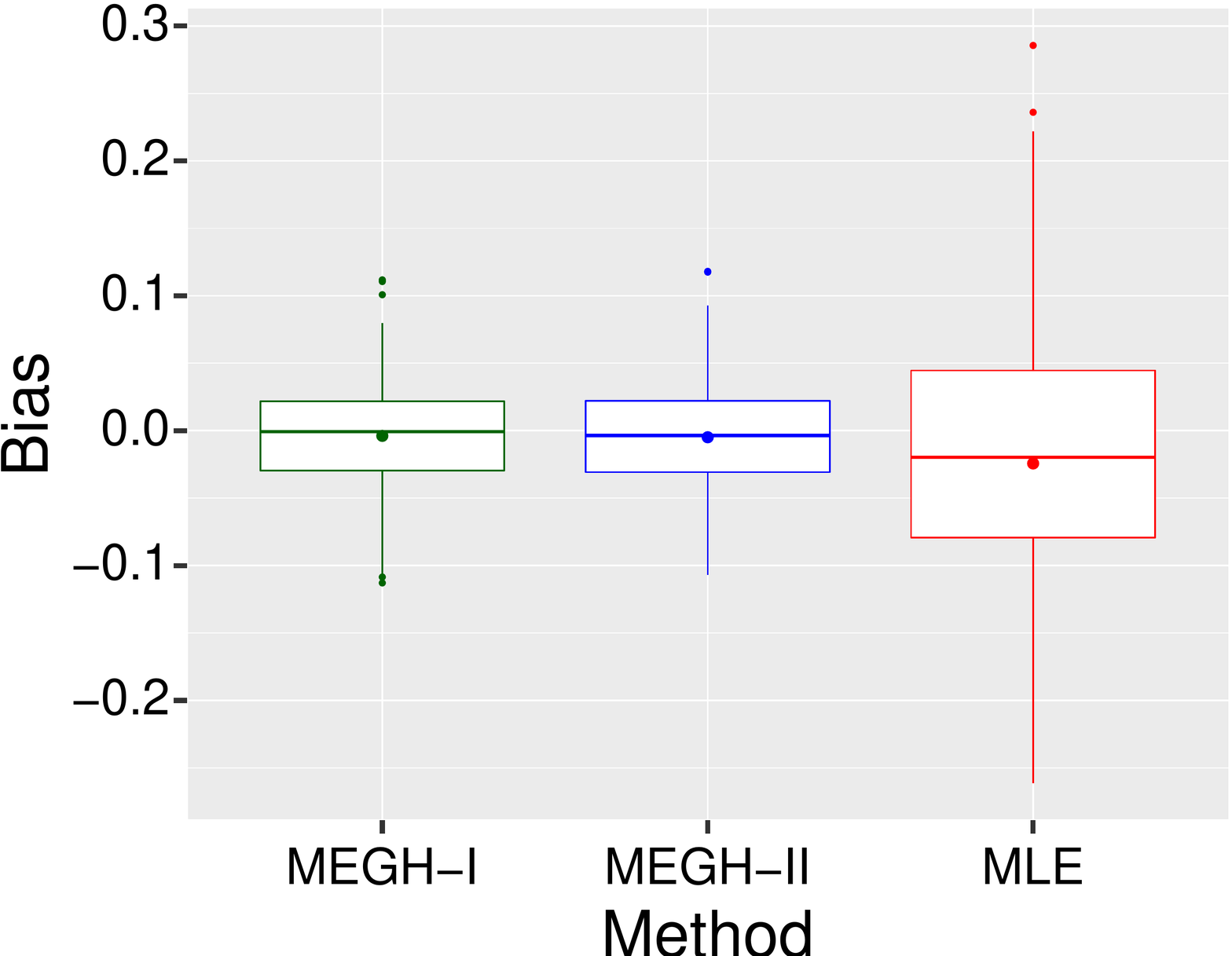}}	\hspace{0.6cm}
	\subfigure[Bias in estimating $\sigma_u$]{\includegraphics[scale=0.17]{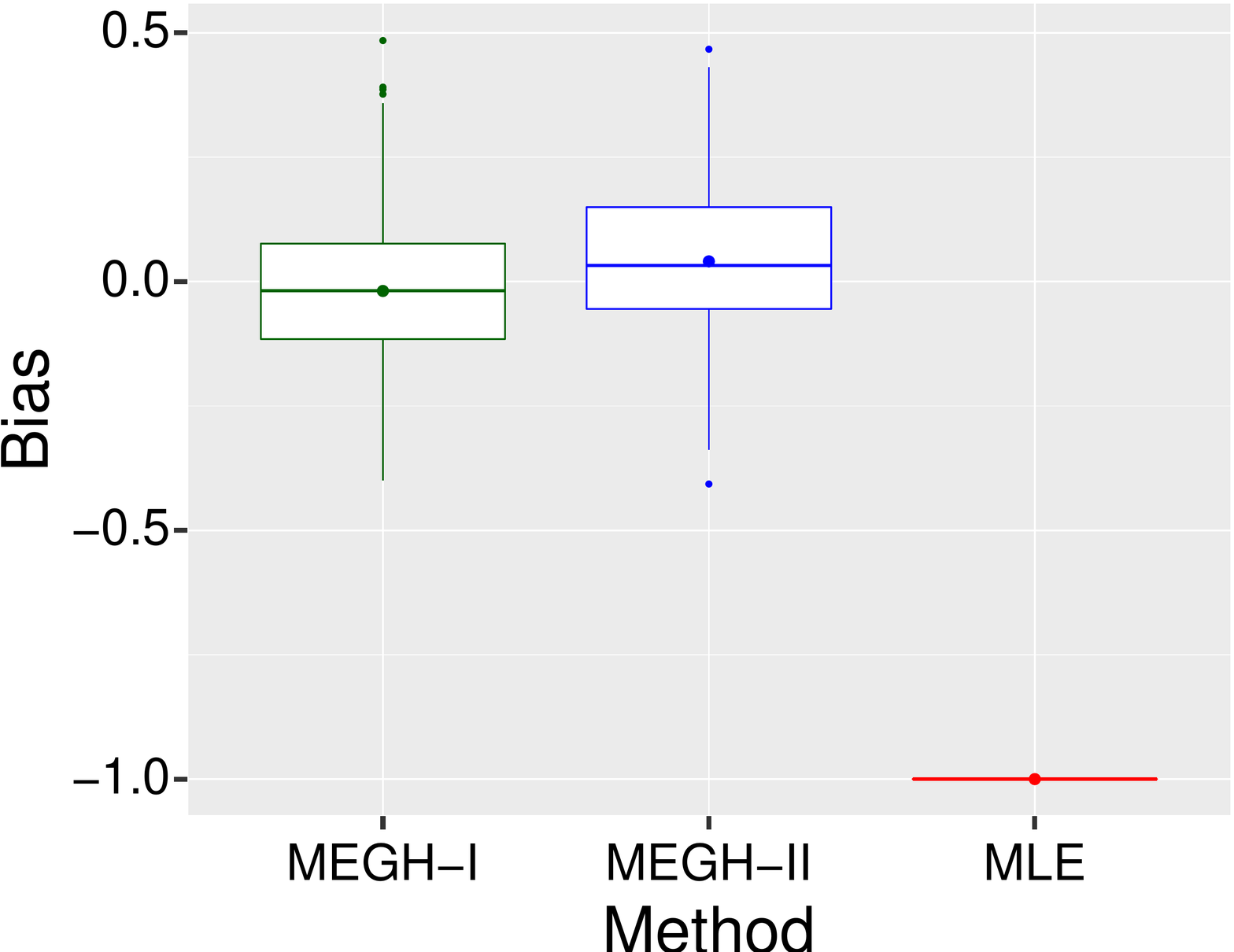}}
\end{figure}

\begin{figure}
	\centering
	\caption{The bias of the estimates from the three methods: MEGH-I, MEGH-II and MLE, for all the parameters based on $250$ simulation replications when the simulated data are generated from model (\ref{s1}) with the mixed structure II and PGW baseline hazard, and a normal distribution for the generated random effects with $\sigma_u=1$.}
	\label{PGW-fig2}
	\vspace*{2mm}
	\subfigure[Bias in estimating $\eta$]{\includegraphics[scale=0.17]{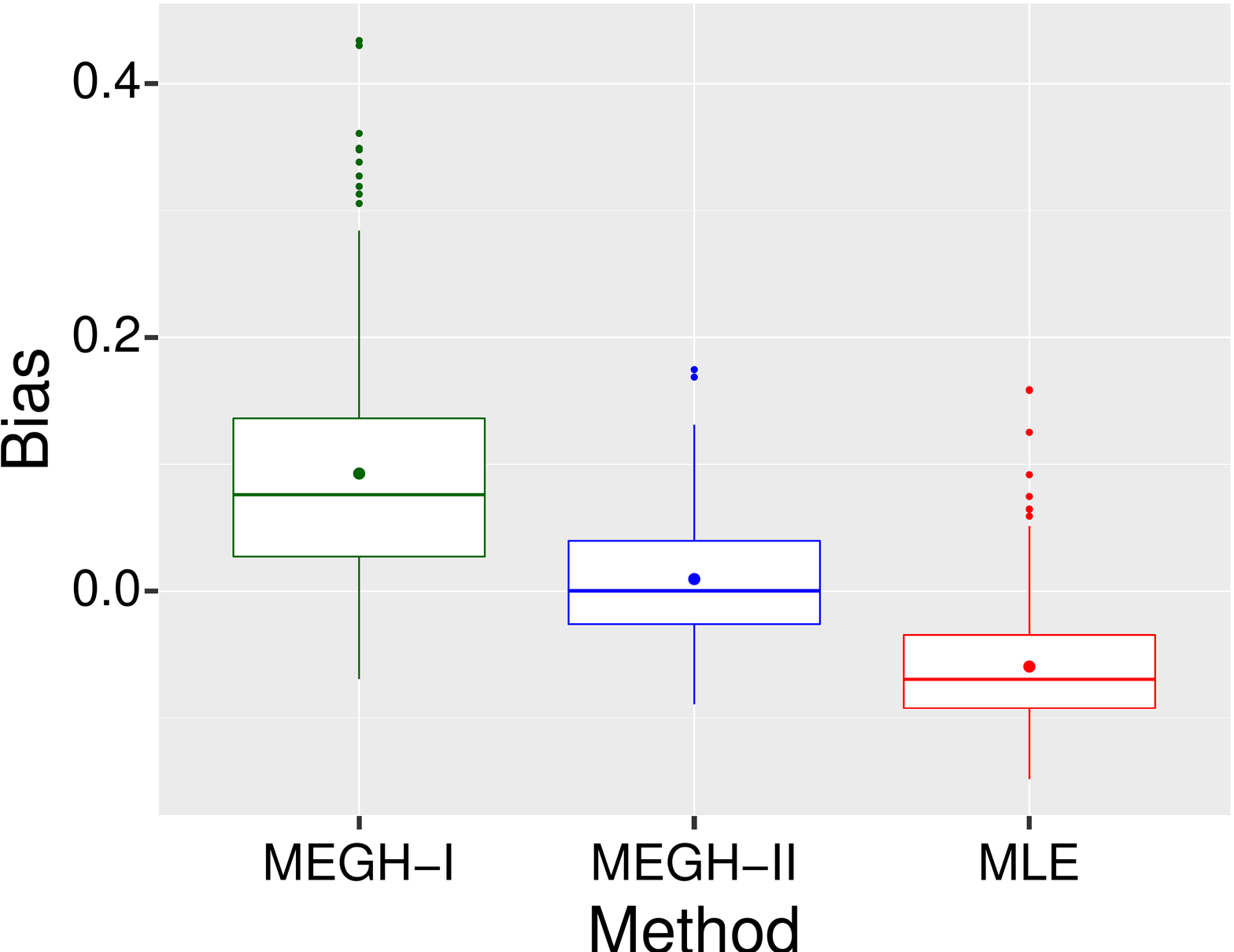}}
	\hspace{0.6cm}
	\subfigure[Bias in estimating $\nu$]{\includegraphics[scale=0.17]{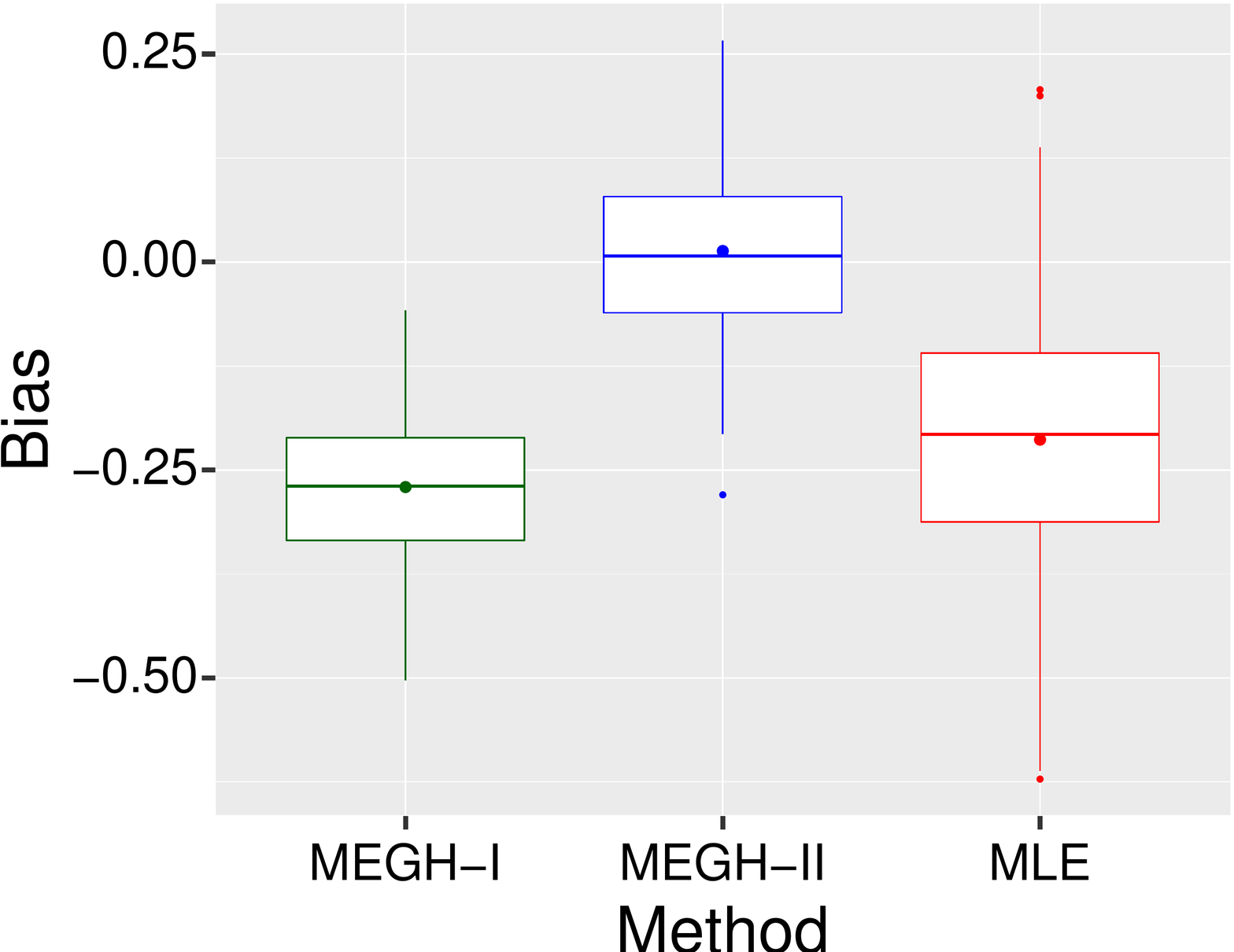}} 
	\hspace{0.6cm}
	\subfigure[Bias in estimating $\delta$]{\includegraphics[scale=0.17]{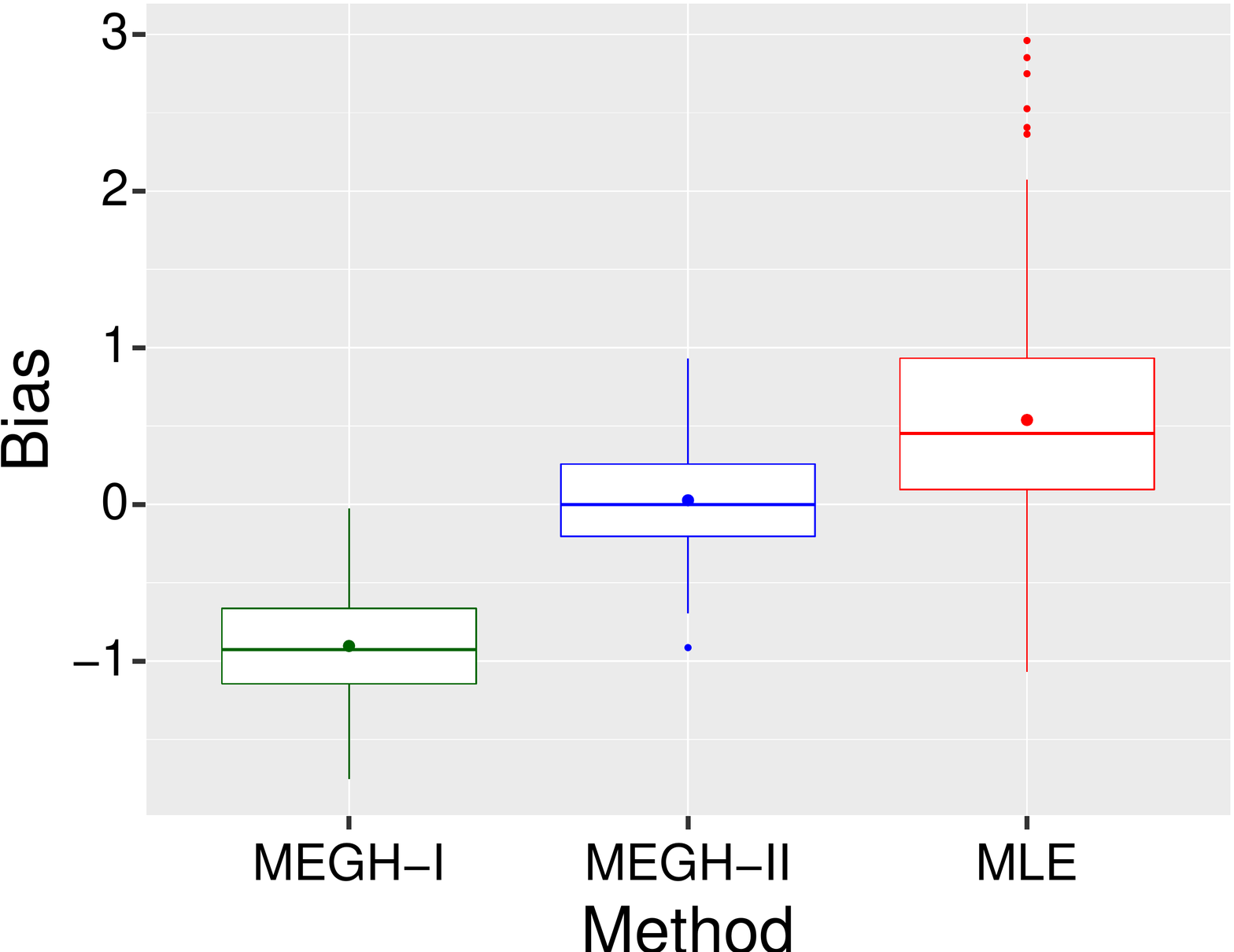}} \\
	\subfigure[Bias in estimating $\alpha$]{\includegraphics[scale=0.17]{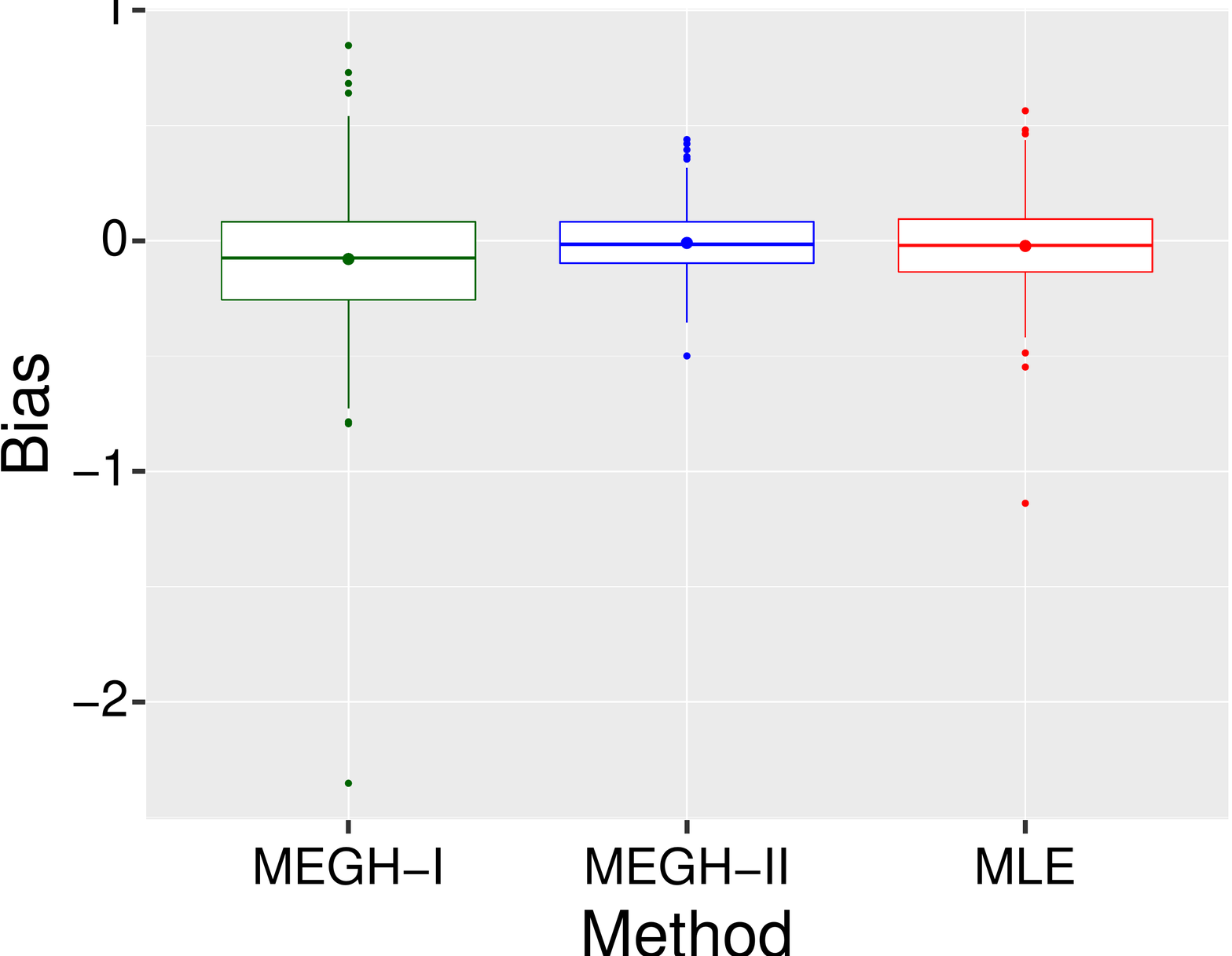}} 
	\hspace{0.6cm}
	\subfigure[Bias in estimating $\beta_1$]{\includegraphics[scale=0.17]{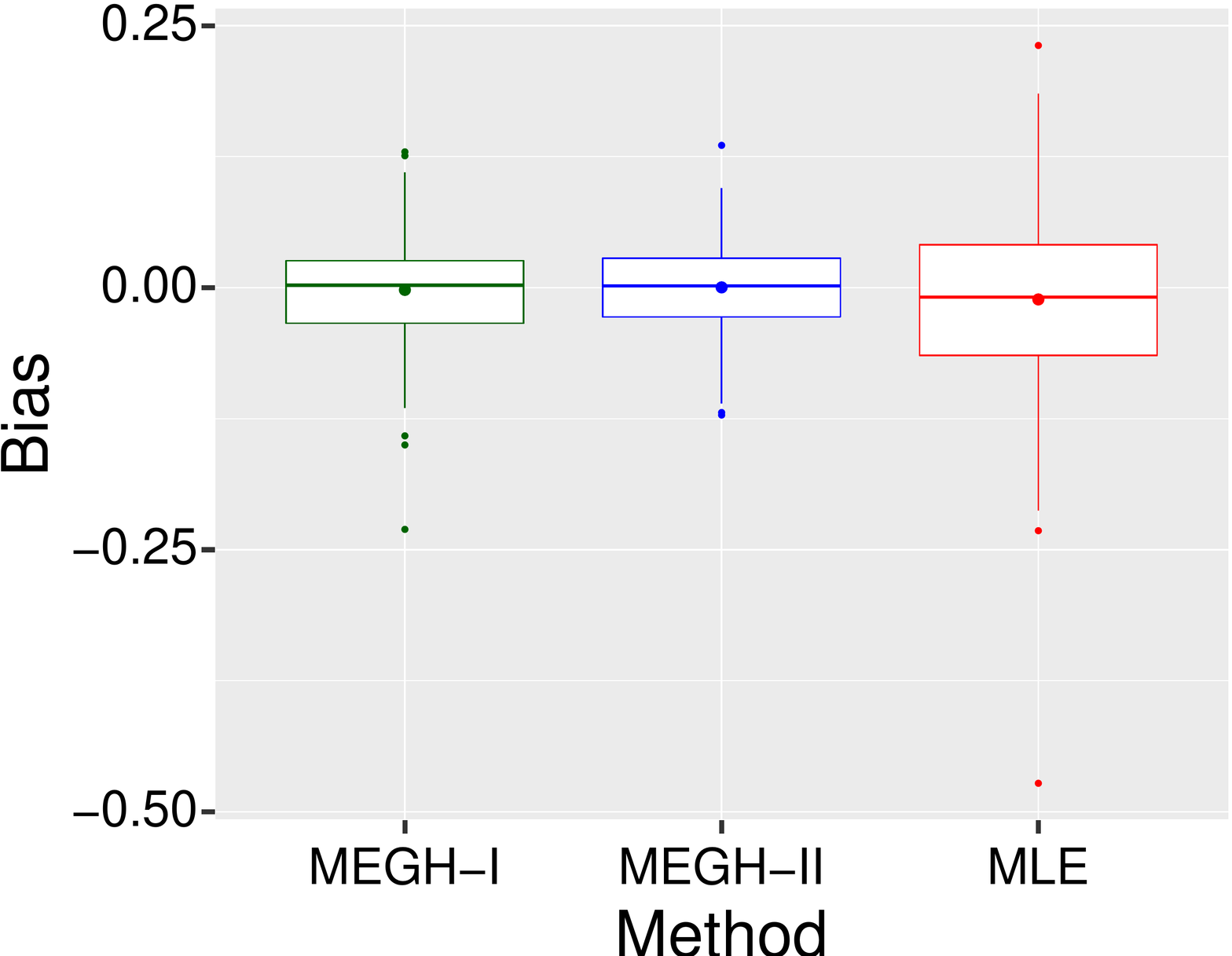}} 
	\hspace{0.6cm}
	\subfigure[Bias in estimating $\beta_2$]{\includegraphics[scale=0.17]{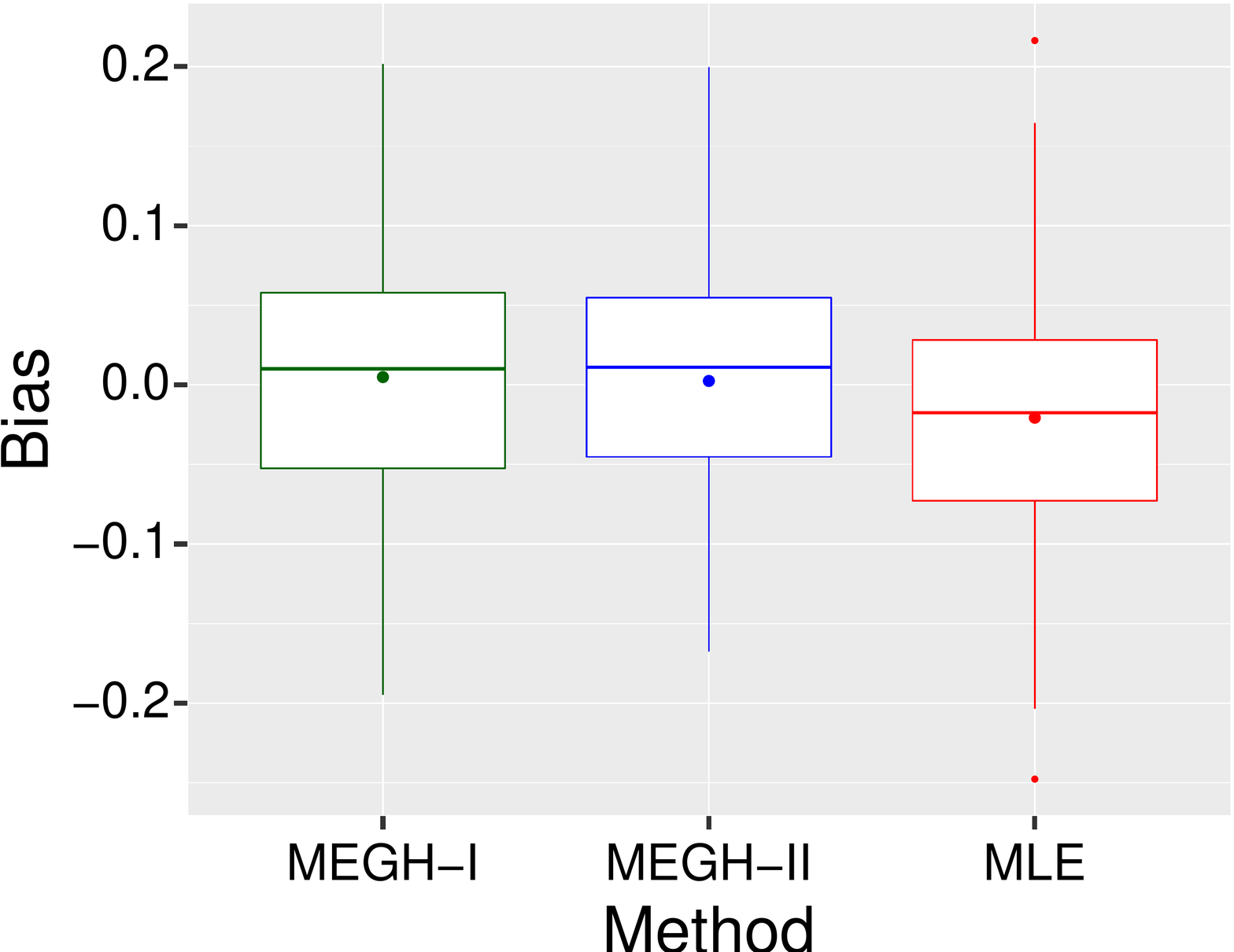}} \\
	\subfigure[Bias in estimating $\beta_3$]{\includegraphics[scale=0.17]{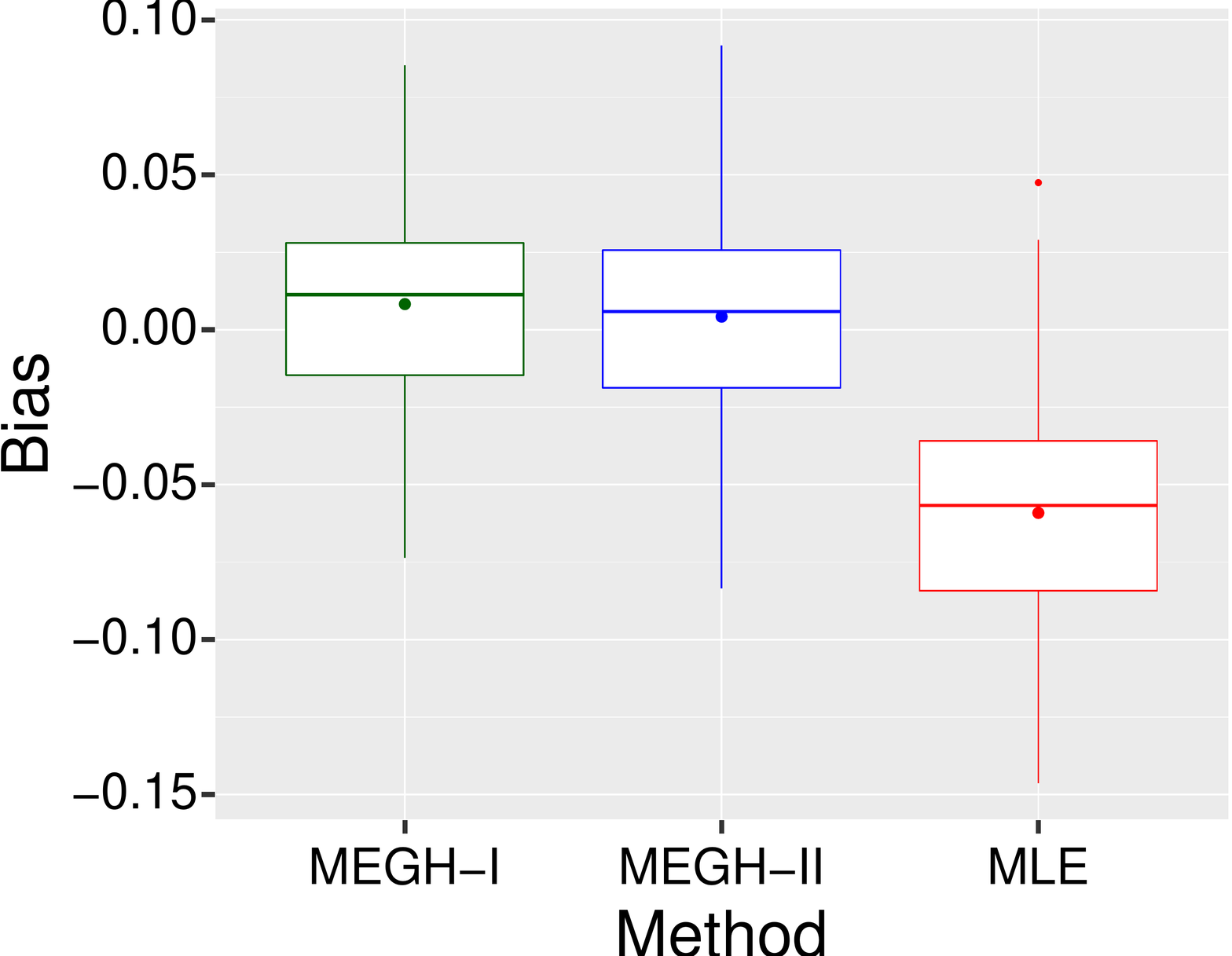}}
	\hspace{0.6cm}
	\subfigure[Bias in estimating $\beta_4$]{\includegraphics[scale=0.17]{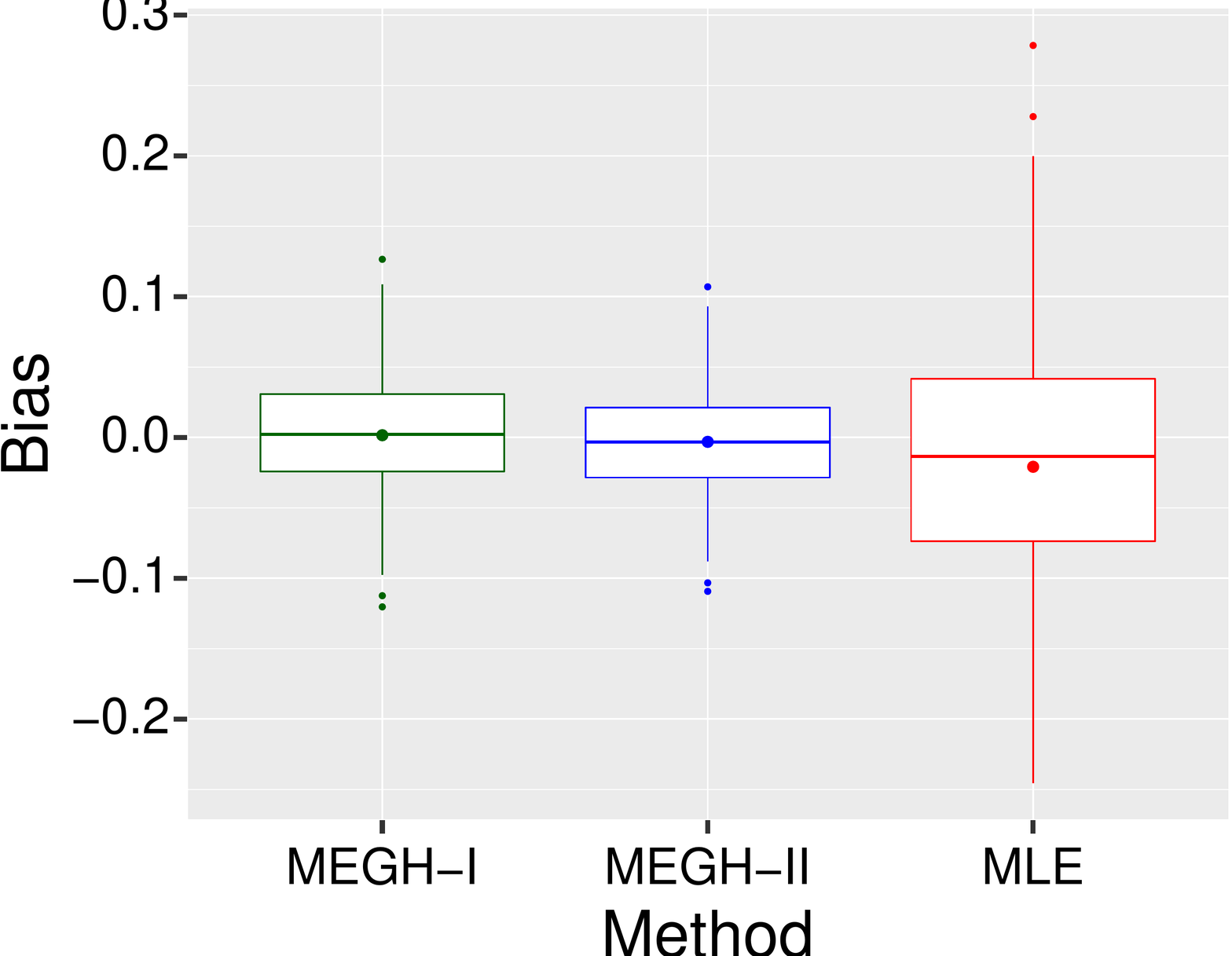}}	
	\hspace{0.6cm}
	\subfigure[Bias in estimating $\sigma_u$]{\includegraphics[scale=0.17]{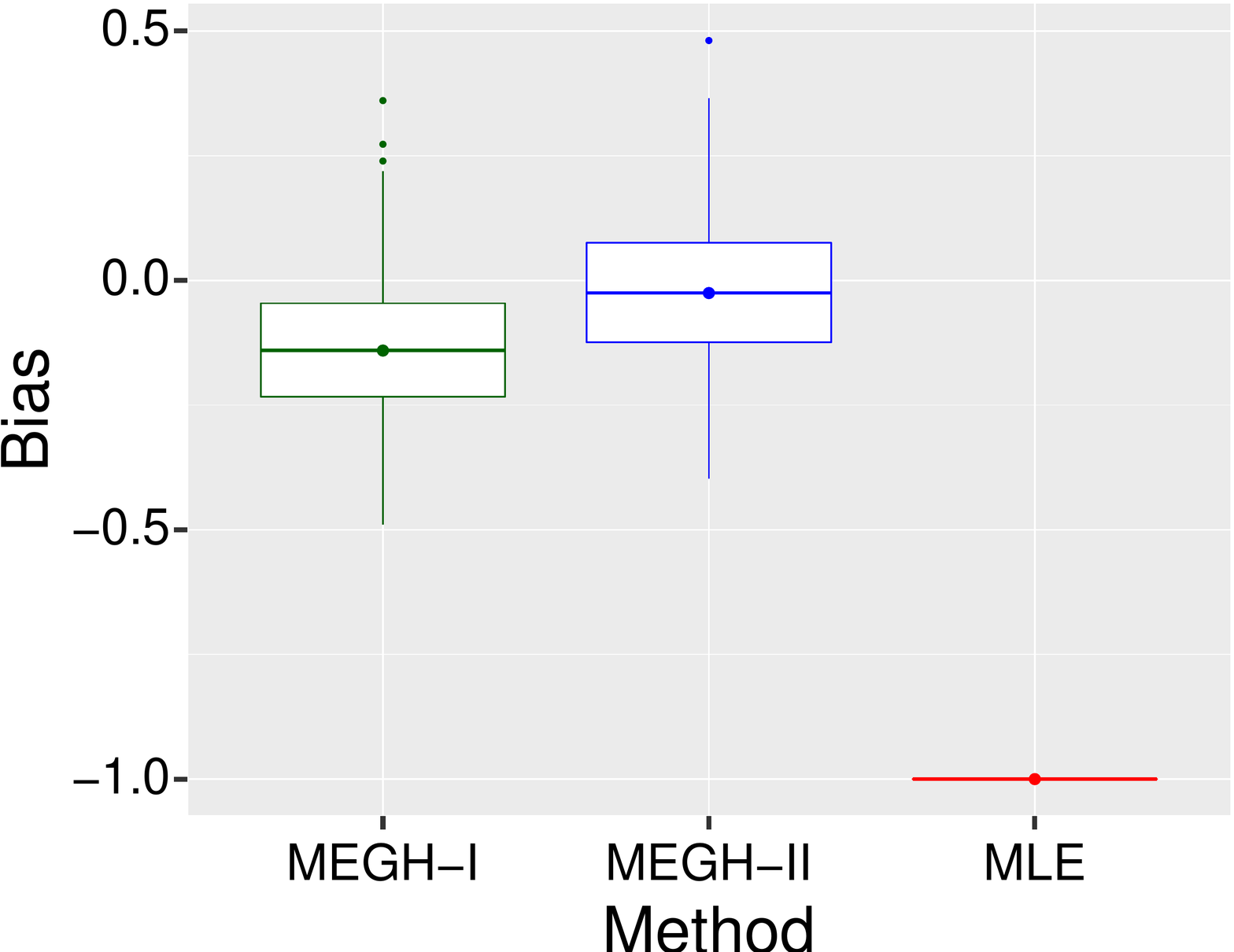}}
\end{figure}

We fit the MEGH model with both mixed hazard structures \eqref{eq:MEGH1}-\eqref{eq:MEGH2} separately to the 250 simulated data sets. We also fit the model ignoring random effects to the simulated data sets. We calculate the bias of the parameter estimates from each of these three models. Figures \ref{PGW-fig1} and \ref{PGW-fig2} show the estimation bias for these three models. It can be seen that both MEGH models with mixed hazard structures \eqref{eq:MEGH1}-\eqref{eq:MEGH2} produce estimates with considerably smaller bias compared to the model ignoring random effects. The bias of the model ignoring random effects for most parameters is higher due to neglecting the presence of random effects. However, we note that the estimates obtained from models with or without random effects are not generally comparable like with like, as discussed in \cite{LeeNelder:2004}. This is because the model with random effects is conditional on random effects, while the model without random effects is marginal (or population-average).

We also observe from the simulation results presented in Figures \ref{PGW-fig1} and \ref{PGW-fig2} that when the mixed hazard structure is misspecified it leads to relatively higher bias compared to the model with the correct mixed hazard structure. This suggests that the choice of the mixed hazard structure is crucial for achieving a reliable model fit and results. We note that the bias of the MEGH model with the correct mixed hazard structure is very small in the simulations.

Table \ref{CIs} presents the $95\%$ confidence intervals for all the regression parameters obtained from the three models considered for the simulation case when the true mixed structure is MEGH-I. The results show that the model with the correct mixed structure MEGH-I produces reliable confidence intervals, but both the model ignoring random effects and the model with the incorrect mixed structure MEGH-II tend to produce inaccurate confidence intervals. In particular, the confidence intervals from the model ignoring random effects mostly do not cover the true parameter values. 

Table \ref{AIC_Power} shows the average AIC of the model fit across the $250$ simulation replications for the three models considered. The results indicate that the MEGH model with the correct mixed hazard structure has the smallest AIC, while the model ignoring random effects tends to produce a very large AIC. Table \ref{AIC_Power} also presents the average power of the likelihood ratio test for random effects across the $250$ simulation replications, when the random effects are generated from a normal distribution with $\sigma_u=1$. It can be seen that the test provides a high power with this sample size. Further simulation results on the performance of the test for random effects under different number of clusters, censoring rates and variance values can be found in the supplementary material.

\begin{table}
	\centering
	\caption{The $95\%$ confidence intervals for all the regression parameters from the three methods: MEGH-I, MEGH-II and MLE, for the case when the true mixed structure is MEGH-I and the random effects are generated from a normal distribution with $\sigma_u=1$. \vspace{-0.2cm}
	\label{CIs}}
	\vspace*{4mm}
	{\tabcolsep=2.0pt
		\begin{tabular}{|c|c|l|c|}
			\hline
			Parameter & True value & Method   & $95\%$ confidence interval  \\	\hline
			&         &  MLE      & $(0.0880, 0.0933)$     \\
			$\eta$     & $0.20$   & MEGH-I    & $(0.1999, 0.2089)$   \\ 
			&         & MEGH-II   & $(0.2815, 0.3017)$     \\ \hline
			&         &  MLE      & $(1.6734, 1.7146)$     \\
			$\nu$    & $1.50$   & MEGH-I    & $(1.5012, 1.5244)$   \\ 
			&         & MEGH-II   & $(1.3492, 1.3762)$     \\ \hline
			&         &  MLE      & $(5.4804, 5.7141)$     \\
			$\delta$   & $3.00$   & MEGH-I    & $(2.9988, 3.0882)$   \\ 
			&         & MEGH-II   & $(2.3658, 2.4574)$     \\ \hline
			&         &  MLE      & $(0.9812, 1.0106)$     \\
			$\alpha$   & $0.96$   & MEGH-I    & $(0.9307, 0.9643)$   \\ 
			&         & MEGH-II   & $(1.1294, 1.1829)$     \\ \hline
			&         &  MLE      & $(0.9847, 1.0010)$     \\
			$\beta_1$   & $1.00$   & MEGH-I    & $(0.9951, 1.0054)$   \\ 
			&         & MEGH-II   & $(0.9676, 0.9782)$     \\ \hline
			&         &  MLE      & $(0.0596, 0.0785)$     \\
			$\beta_2$   & $0.08$   & MEGH-I    & $(0.0741, 0.0919)$   \\ 
			&         & MEGH-II   & $(0.0713, 0.0891)$     \\ \hline
			&         &  MLE      & $(0.1540, 0.1633)$     \\
			$\beta_3$   & $0.22$   & MEGH-I    & $(0.2204, 0.2289)$   \\ 
			&         & MEGH-II   & $(0.2172, 0.2258)$     \\ \hline
			&         &  MLE      & $(0.0641, 0.0872)$     \\
			$\beta_4$    & $0.10$   & MEGH-I    & $(0.0914, 0.1009)$   \\ 
			&         & MEGH-II   & $(0.0901, 0.1001)$     \\ \hline
	\end{tabular}}
\end{table}

\begin{table}
	\centering
	\caption{Average AIC of the model fit and average power of the test for random effects across $250$ simulation replications, when the random effects are generated from a normal distribution with $\sigma_u=1$. \vspace{-0.2cm}  
	\label{AIC_Power}}
	\vspace*{4mm}
	{\tabcolsep=2.0pt
		\begin{tabular}{|l|l|cc|}
			\hline
			True structure & Fitted model & AIC    & Power  \\	\hline
			&  MLE      &$1390.66$   &$-$     \\
			MEGH-I     &  MEGH-I   &$927.84$   &$1.0$   \\ 
			&  MEGH-II  &$975.79$   &$1.0$     \\ \hline
			&  MLE      &$1383.18$    &$-$     \\
			MEGH-II    &  MEGH-I   &$996.35$    &$1.0$   \\  
			&  MEGH-II  &$962.57$    &$1.0$   \\ \hline
	\end{tabular}}
\end{table}

\section{Real Data Application}
\label{realdatasection}
\begin{figure}
	\centering
	\caption{Leukemia data: Kaplan-Meier estimators of survival by district.}
	\label{fig:KMs}
	\vspace*{2mm}
	\includegraphics[scale=0.65]{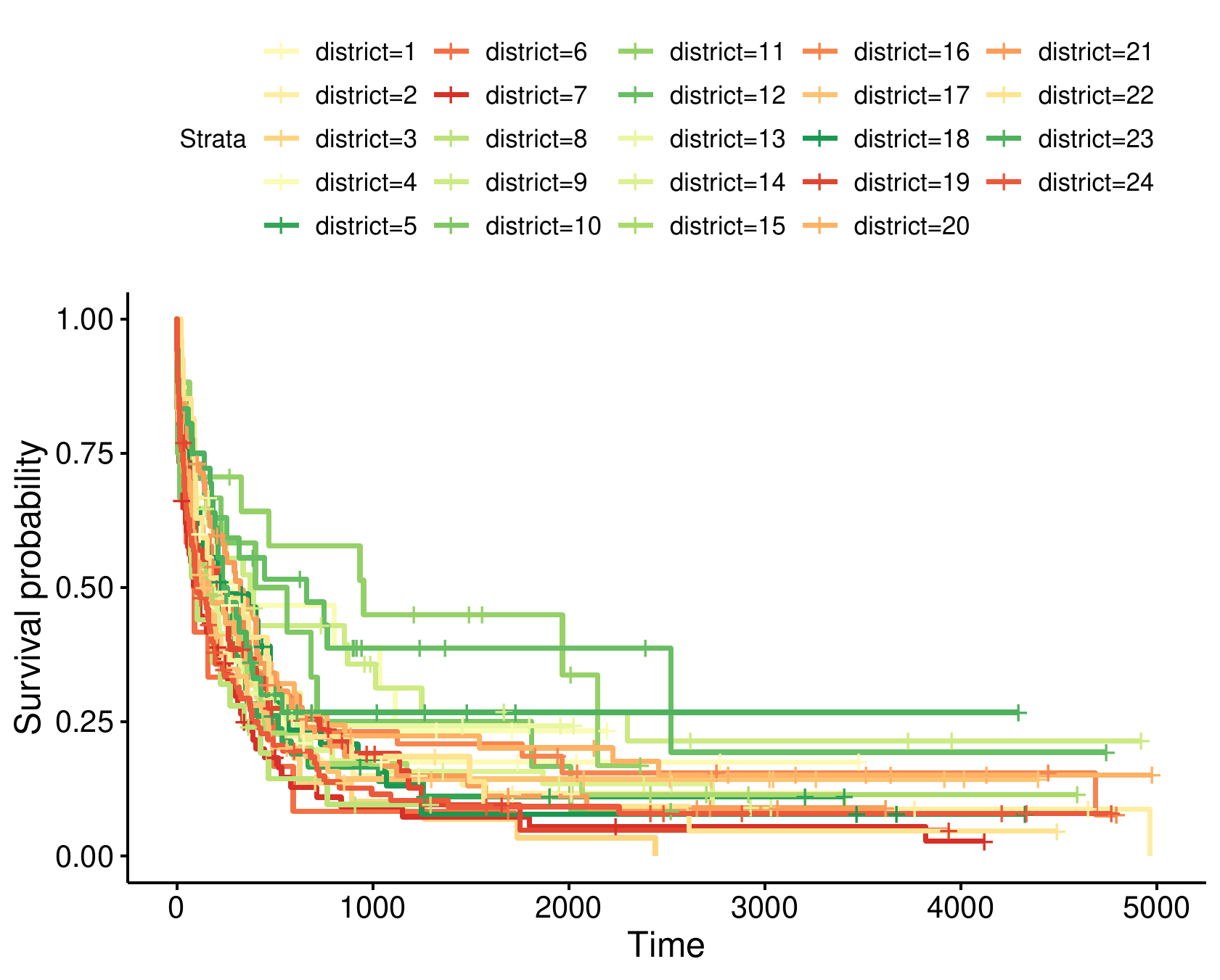}
\end{figure}

In this section, we apply the MEGH model to analyse the data set {\tt LeukSurv}, which contains information on the survival of $n = 1043$ patients of acute myeloid leukemia in northwest England, recorded between 1982 and 1998. 
This data set is available in the R package {\tt spBayesSurv} \cite{zhou:2018}. Previous analyses of this data set have suggested that there is evidence of variation in survival across this region \cite{zhou:2018}, and we can see that this is also suggested by the nonparametric Kaplan-Meier estimators of the survival curves by district shown in Figure \ref{fig:KMs}.
We use information about the survival time (in years), vital status at the end of follow-up (0 - right-censored, 1 - dead), age (in years), sex (0 - female, 1 - male), white blood cell count at diagnosis ({\tt wbc}, truncated at 500), the Townsend score ({\tt tpi}, higher values indicates less affluent areas), and administrative district of residence ({\tt district}, 24 districts). We fit the MEGH models with hazard structures MEGH-I \eqref{eq:MEGH1} and MEGH-II \eqref{eq:MEGH2}, and with the PGW and log-logistic baseline hazards. For the random-effects distribution, we use the normal, Student-$t$ and two-piece normal distributions (to account for heavier tails than normal and asymmetry).
In all fitted models, we consider the time dependent effect ($\tilde{\bx}_{ij}$) of age (standardised), and the hazard-level effects ($\bx_{ij}$) of age (standardised), sex, white blood cell count at diagnosis ({\tt wbc}, standardised) and the Townsend score ({\tt tpi}, standardised). The variable {\tt district} is used to define the random effects in the two mixed hazard structures considered. In addition, we fit the models ignoring random effects for comparison. We also assess the need for including random effects using the diagnostic tests in Section \ref{sec_testsRE}.

\begin{figure}
	\centering
	\caption{Leukemia data: maps showing the aggregated summaries for each district: (a) mean age, (b) mean Townsend score, (c) marginal 1-year survival, and (d) marginal 5-year survival.}
	\label{fig:maps}
	\subfigure[]{\includegraphics[scale=0.37]{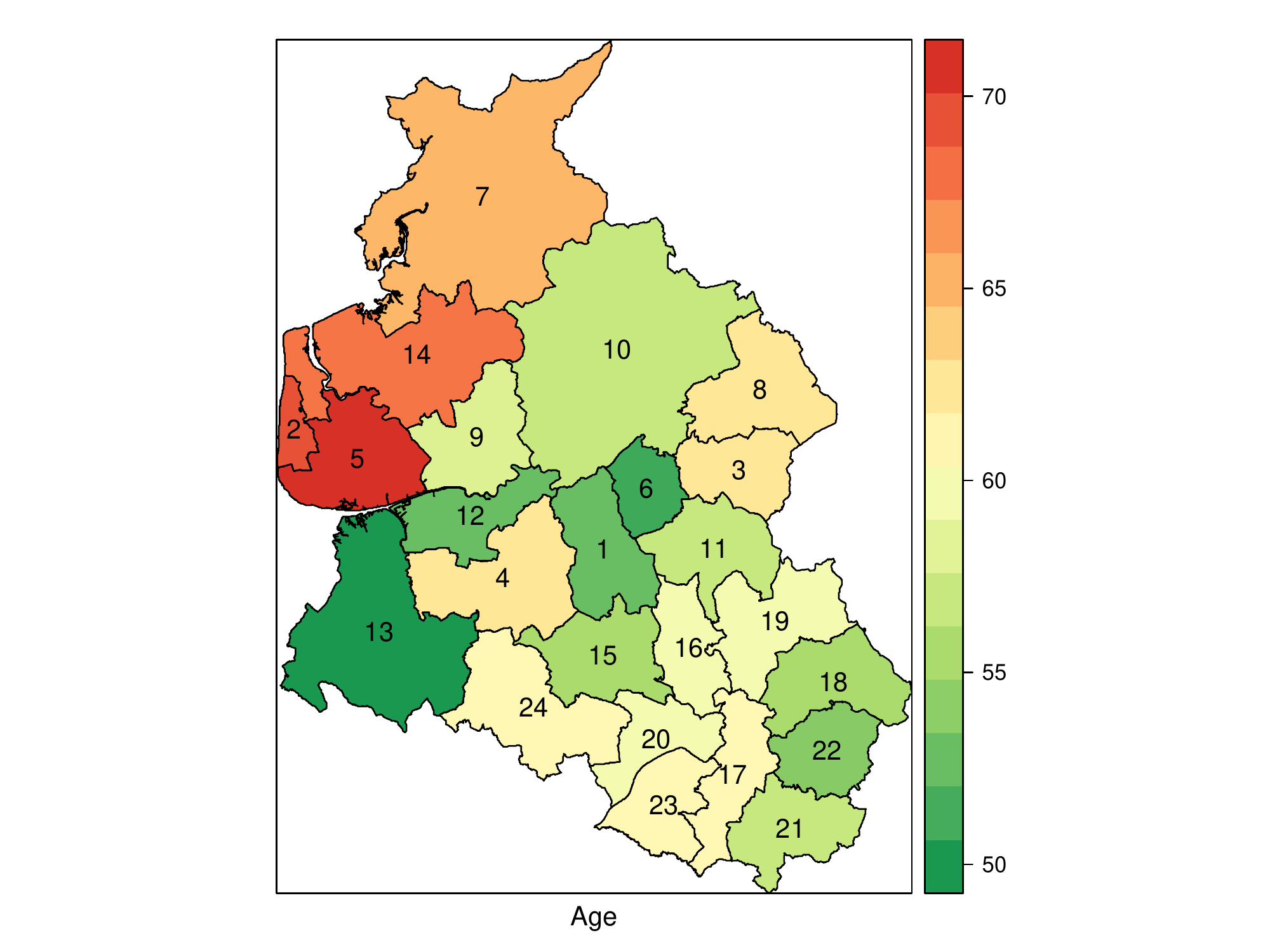}}
	\hspace{0.35cm}
	\subfigure[]{\includegraphics[scale=0.37]{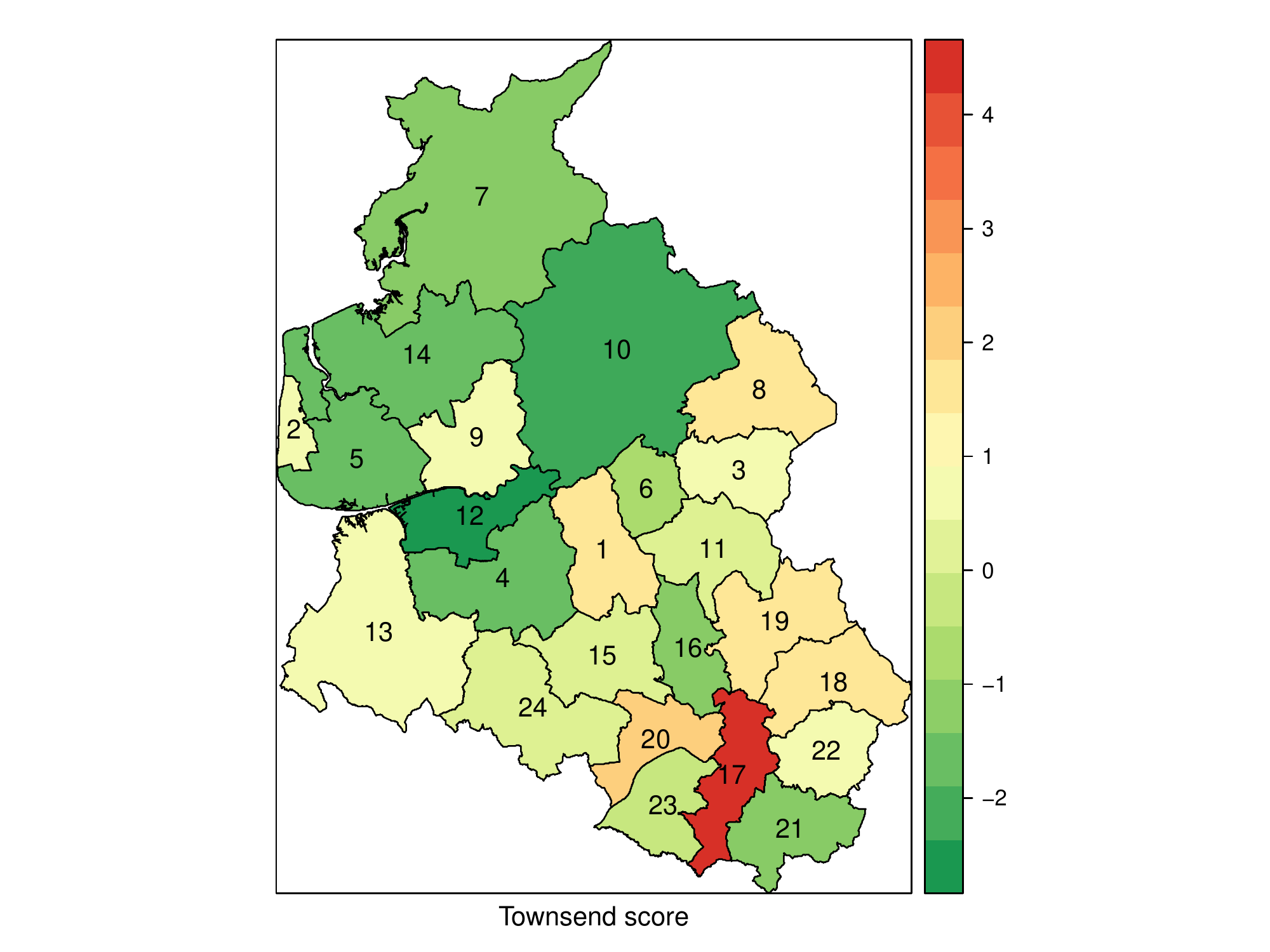}} \\
	\subfigure[]{\includegraphics[scale=0.37]{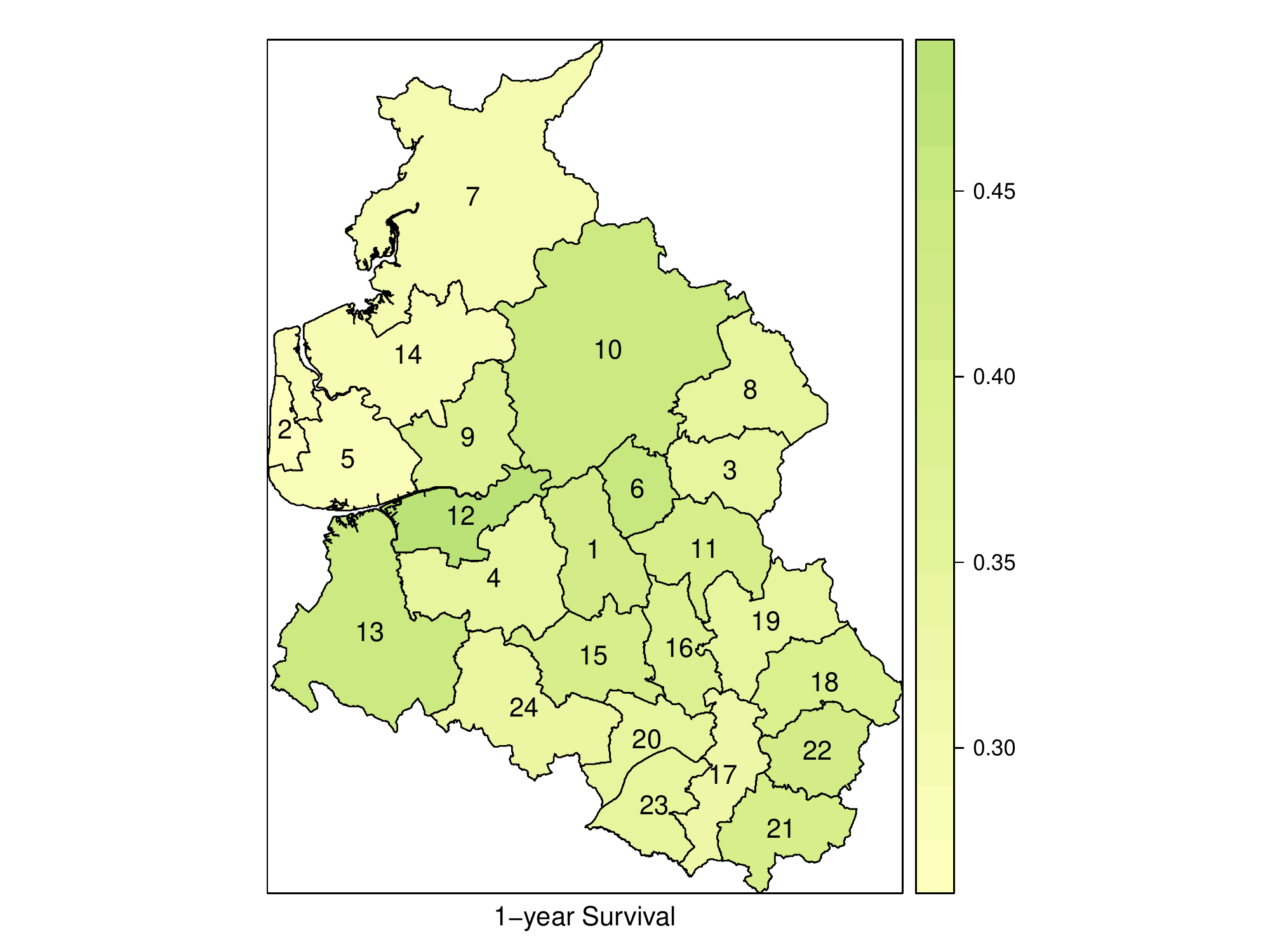}} 
	\hspace{0.35cm}
	\subfigure[]{\includegraphics[scale=0.37]{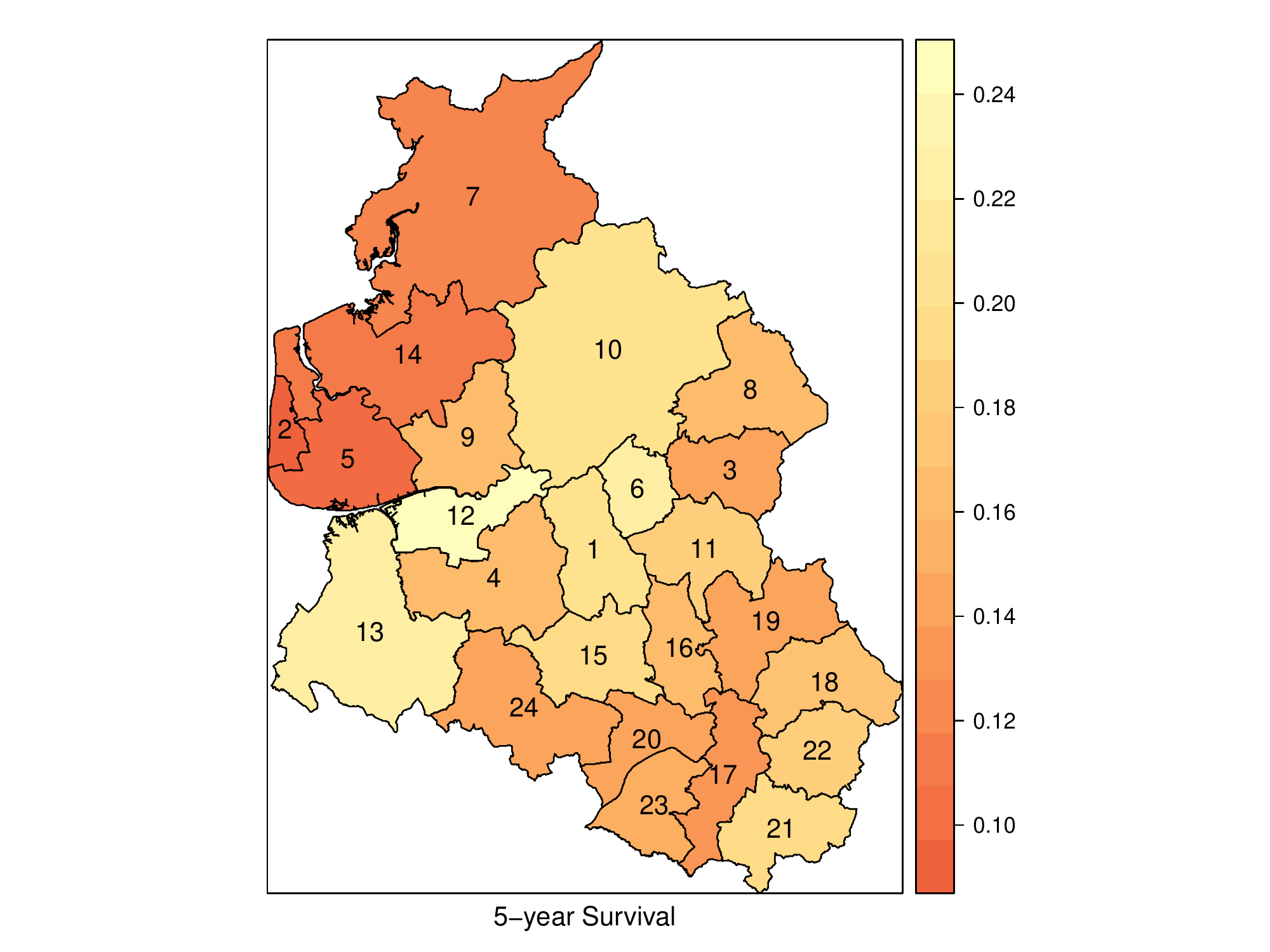}} 
\end{figure}

The best model selected using AIC (1553.725) is the model with the mixed structure MEGH-I, log-logistic baseline hazard (with log-location and scale parameters $(\theta_1,\theta_2) = (\mu,\tau)$) and normal random effects. The AIC for the model ignoring random effects with log-logistic baseline hazard is 1556.366, which is fairly close to the AIC value for the best model. The reason for this is that the estimate of the variance of random effects is relatively small ($\widehat{\sigma}_u = 0.144$), and the number of clusters ($r=24$) is not large in this data set. However, the p-value for testing $H_0:\,\sigma_u = 0$ is $0.0156$, which suggests that the random effect is significant and must be present in the model. This implies that there is significant between-cluster variability after incorporating the information on the observed covariates (see also the Box and Whisker plot in the online Supplementary Material). 
Figure \ref{fig:maps} shows the maps of aggregated summaries at district level. We notice from Figure \ref{fig:maps}c-d that the marginal survival functions (which is obtained as the average of cluster-specific marginal survival functions using Monte Carlo integration) at $t=1,5$ years exhibit some variability by district. Figures \ref{fig:maps}a-b show that the distribution of mean age and mean Townsend score is also quite different for the different districts. Consequently, survival gaps between different districts are explained by a combination of complex factors, including different age and Townsend distribution, a conclusion that could be harder to obtain with fully nonparametric methods that require data stratification. 

\begin{figure}
	\centering
	\caption{Leukemia data: (a) gradient function plot with the $95\%$ confidence bands 
	 for the MEGH model \eqref{eq:MEGH1} with normal random effects fitted to the leukemia data.
		(b) gradient function plot with the $95\%$ confidence bands for the MEGH model \eqref{eq:MEGH1} with a two-piece normal distribution for the random-effects.}
	\label{figure_gradient}
	\vspace*{2mm}
	\subfigure[]{\includegraphics[scale=0.25]{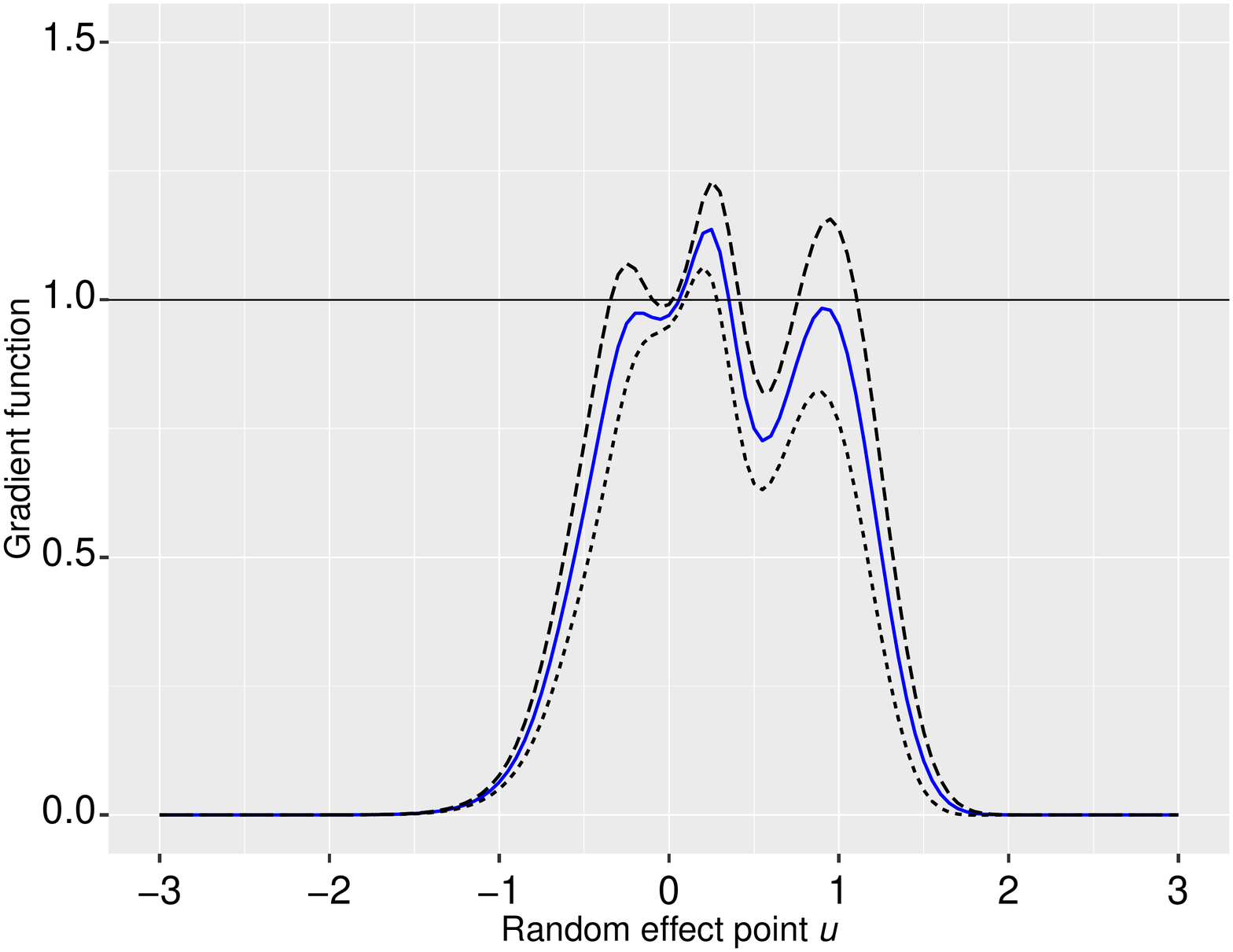}}
	\hspace{0.7cm}
	\subfigure[]{\includegraphics[scale=0.25]{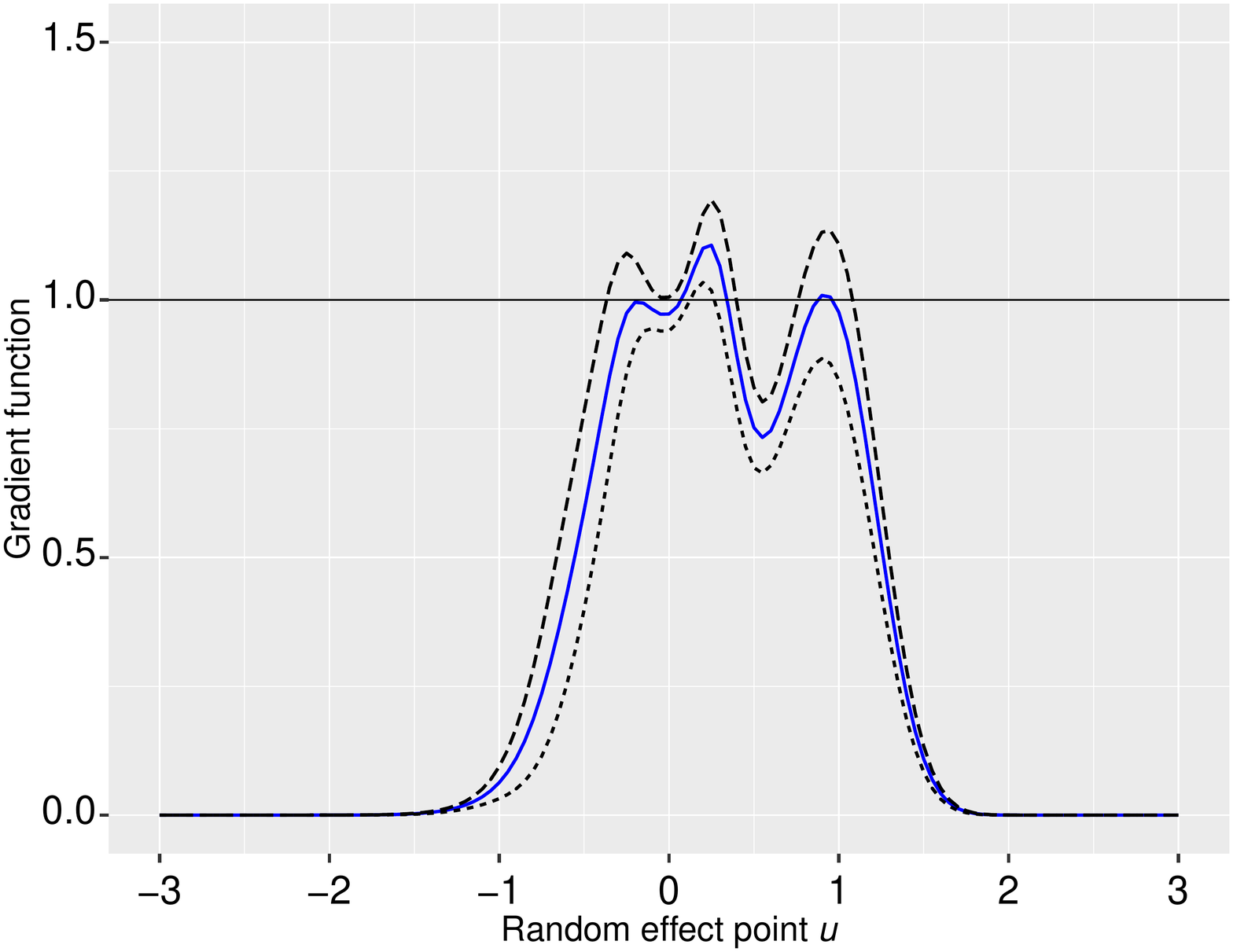}}
\end{figure}

The gradient function plot for the model MEGH-I fitted under normal random effects is presented in Figure \ref{figure_gradient}a, together with the $95\%$ confidence bands similar to \cite{verbeke2013}, which indicates that the normality assumption on random effects is valid since the gradient values remain under or close to $1$. To double check whether the fit can be improved by using a distribution for random effects which allows to capture asymmetry, we calculate the gradient function plot with the two-piece normal distribution for random effects. The gradient plot with the $95\%$ confidence bands for the two-piece normal distribution, shown in Figure \ref{figure_gradient}b, is very similar, suggesting that the usual normal distribution is adequate for random effects.

\section{Discussion}\label{sec:discussion}
We have introduced a general mixed-effects hazard structure (MEGH), as well as two general subclasses of interest (MEGH -I and MEGH-II). The proposed structures generalise classical survival regression models of interest in practice, such as the MEPH and MEAFT. We have adopted a flexible parametric approach, which allowed us to prove asymptotic results under standard regularity conditions. 
Our simulation studies show that the estimates of the parameters of the proposed models, based on the marginal likelihood, have good frequentist properties.
Another contribution of this work consists of evaluating the impact of misspecifying the role of the random effects and potentially the random effects distribution.  Our simulation study shows that not only misspecifying the distribution of the random effects has an impact on the quality of the inference on the parameters, but also misspecifying the role of the random effects in the hazard regression model, a topic that has received little attention in survival analysis. 
The routines used to produce our results are implemented in the R package {\tt MEGH} which is available, along with the real data application, at \url{https://github.com/FJRubio67/MEGH}. In our simulations and applications, we have focused on the use of the MEGH-I and MEGH-II structures, as these represent generalisations of the most common mixed models in survival analysis. However, the implementation of the general structure \eqref{eq:MEGH}, with two random effects, represents a possible extension of this work. Such extension does not only bring computational challenges, as one requires double numerical integration, but also the modelling of the dependence between the two random effects. The latter can be explored by using copulas to capture different types of dependencies, and a variety of parametric marginal distributions \cite{rubio:2018b}

The tractability of the MEGH model allows for several extensions, such as accounting for left-censored or interval censored times. A more substantial extension corresponds to the case where the distribution of the random effects reflects a spatial multilevel structure \cite{zhou:2020}. Another possible extension consists of modelling the baseline hazard using nonparametric (frequentist or Bayesian) methods \cite{chen:2001}. Extensions to the inclusion of more than one hierarchical level would also be of practical interest. This could be done simply by considering multivariate random effects:
\begin{equation*}
h(t_{ij}\mid \bx_{ij}, \bu_i, \tilde{\bu}_i) = h_0\left(t_{ij} \exp\left\{\tilde{\bx}_{ij}^{\top}\balpha + \tilde{\bz}_{ij}^{\top}\tilde{\bu}_i \right\}\right)\exp\left\{\bx_{ij}^{\top}\bbeta + \bz_{ij}^{\top}\bu_i \right\},
\end{equation*}
where $\bz_{ij}$ and $\tilde{\bz}_{ij}$ are random effects covariates associated with the hazard and time scales respectively, and $(\bu_i,\tilde{\bu}_i) \stackrel{iid}{\sim} G_{q+\tilde{q}}$, where $G_{q+\tilde{q}}$ is a continuous distribution with support on ${\mathbb R}^{q + \tilde{q}}$ and zero mean.
We emphasise that, although interesting, this strategy requires a careful study of the identifiability of parameters.
Our formulation remains valid for these extensions, but the main challenges relate to the development of computational methods for addressing those questions.

\bibliographystyle{vancouver}

\end{document}